\documentclass[twocolumn,compsoc,journal]{IEEEtran}
\usepackage[T1]{fontenc}
\usepackage[latin9]{inputenc}
\usepackage{array}
\usepackage{float}
\usepackage{booktabs}
\usepackage{amsmath}
\usepackage{amssymb}
\usepackage{graphicx}
\usepackage[unicode=true,
 bookmarks=true,bookmarksnumbered=true,bookmarksopen=true,bookmarksopenlevel=1,
 breaklinks=false,pdfborder={0 0 0},pdfborderstyle={},backref=false,colorlinks=false]
 {hyperref}
\hypersetup{pdftitle={Your Title},
 pdfauthor={Your Name},
 pdfpagelayout=OneColumn, pdfnewwindow=true, pdfstartview=XYZ, plainpages=false}

\makeatletter

\providecommand{\tabularnewline}{\\}
\floatstyle{ruled}
\newfloat{algorithm}{tbp}{loa}
\providecommand{\algorithmname}{Algorithm}
\floatname{algorithm}{\protect\algorithmname}

\let\oldforeign@language\foreign@language
\DeclareRobustCommand{\foreign@language}[1]{%
  \lowercase{\oldforeign@language{#1}}}

\usepackage[caption=false,font=normalsize,labelfont=sf,textfont=sf]{subfig}
\usepackage{algorithm}
\usepackage{algpseudocode}
\algnewcommand{\Linecomment}[1]{\Statex \(\triangleright\) #1}
\algrenewcommand\algorithmicrequire{\textbf{Input:}}
\algrenewcommand\algorithmicensure{\textbf{Output:}}
\usepackage[justification=centering]{caption}

\makeatother

\begin{document}
\title{A Scalable Deep Reinforcement Learning Model for Online Scheduling
Coflows of Multi-Stage Jobs for High Performance Computing}
\author{Xin~Wang and Hong~Shen~\IEEEcompsocitemizethanks{\IEEEcompsocthanksitem Authors are with School of Computer Science
and Engineering, Sun Yat-sen University, China. Corresponding author
is Hong Shen.\protect \\
E-mail: \href{http://shenh3@mail.sysu.edu.cn}{shenh3@mail.sysu.edu.cn}.}\thanks{Manuscript received April 19, 2005; revised December 27, 2012.}}
\markboth{IEEE TRANSACTIONS ON PARALLEL AND DISTRIBUTED SYSTEMS}{ Wang \MakeLowercase{\textit{et al.}}: A Scalable Deep Reinforcement
Learning Model for Online Scheduling Coflows of Multi-Stage Jobs for
High Performance Computing}
\IEEEpubid{0000\textendash 0000/00\$00.00~\copyright~2012 IEEE}

\IEEEtitleabstractindextext{
\begin{abstract}
Coflow is a recently proposed networking abstraction to help improve
the communication performance of data-parallel computing jobs. In
multi-stage jobs, each job consists of multiple coflows and is represented
by a Directed Acyclic Graph (DAG). Efficiently scheduling coflows
is critical to improve the data-parallel computing performance in
data centers. Compared with hand-tuned scheduling heuristics, existing
work DeepWeave \cite{literature14} utilizes Reinforcement Learning
(RL) framework to generate highly-efficient coflow scheduling policies
automatically. It employs a graph neural network (GNN) to encode the
job information in a set of embedding vectors, and feeds a flat embedding
vector containing the whole job information to the policy network.
However, this method has poor scalability as it is unable to cope
with jobs represented by DAGs of arbitrary sizes and shapes, which
requires a large policy network for processing a high-dimensional
embedding vector that is difficult to train. In this paper, we first
utilize a directed acyclic graph neural network (DAGNN) to process
the input and propose a novel Pipelined-DAGNN, which can effectively
speed up the feature extraction process of the DAGNN. Next, we feed
the embedding sequence composed of schedulable coflows instead of
a flat embedding of all coflows to the policy network, and output
a priority sequence, which makes the size of the policy network depend
on only the dimension of features instead of the product of dimension
and number of nodes in the job's DAG. Furthermore, to improve the
accuracy of the priority scheduling policy, we incorporate the Self-Attention
Mechanism into a deep RL model to capture the interaction between
different parts of the embedding sequence to make the output priority
scores relevant. Based on this model, we then develop a coflow scheduling
algorithm for online multi-stage jobs. Our simulation results are
based on the real trace of Facebook. Compared with a state-of-the-art
approach, our model can shorten the average weighted job completion
time by up to 40.42$\%$ and complete jobs at least 1.68 times faster.
It also has better scalability and robustness.
\end{abstract}


}
\maketitle

\IEEEdisplaynontitleabstractindextext{}

\IEEEpeerreviewmaketitle{}

\section{Introduction}

\label{sec:introduction}

\IEEEPARstart{D}{ata} parallel frameworks, such as MapReduce \cite{literature1},
Hadoop \cite{literature2} and Spark \cite{literature3} are very
popular in cloud applications. In typical data center applications,
the execution process is an application consisting of multiple consecutive
stages. Each stage relies on several parallel flows, and the next
stage cannot begin until all flows in the current stage have been
transmitted. Therefore, the traditional network metrics, such as minimizing
the average flow completion time (FCT), though can guarantee the performance
at the flow level, is unable to effectively improve the transmission
performance of the application.

\begin{figure}[tbh]
\centering \includegraphics[width=0.3\textwidth,height=3.2cm]{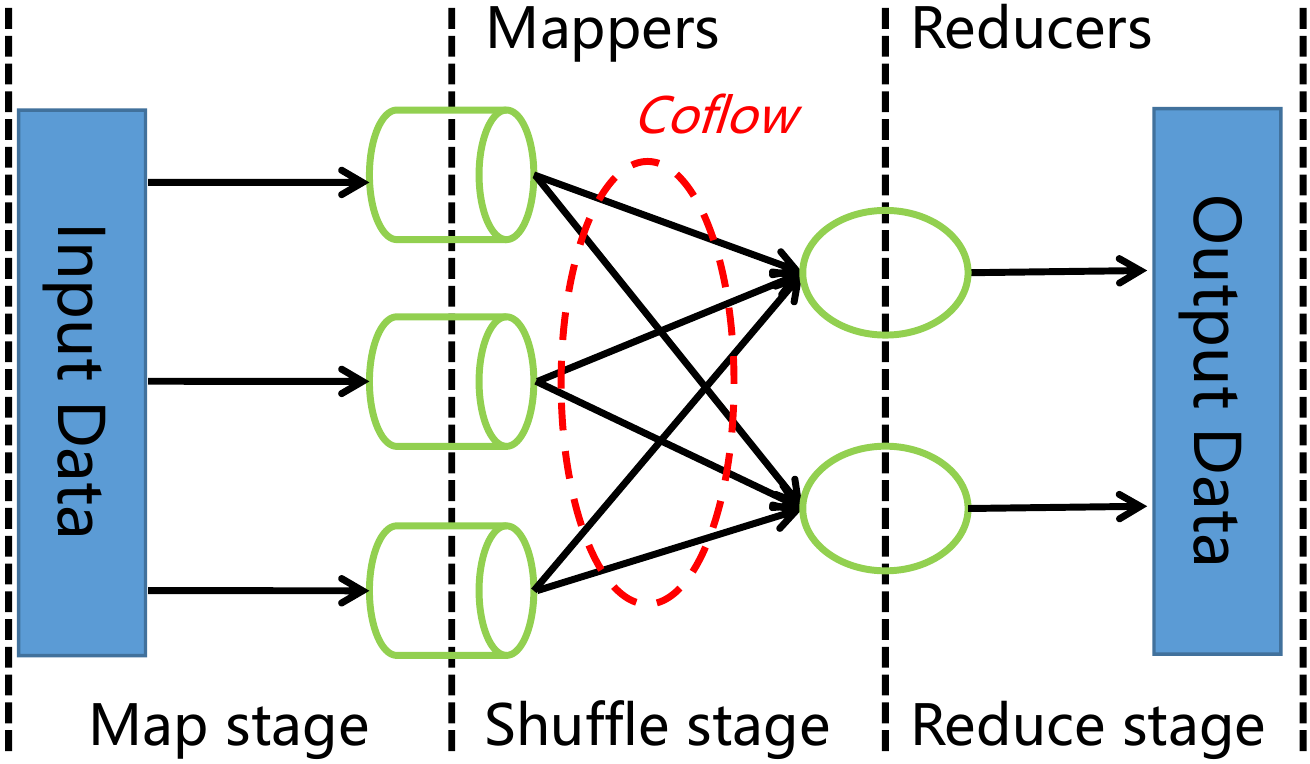}
\caption{Coflow abstraction in parallel computing}
\label{fig:MapReduce}
\end{figure}

The $\textit{coflow}$ abstraction was first proposed in \cite{literature5}
to improve the application-level communication performance, which
is defined as a collection of parallel flows with the same application
semantic, usually appearing between two stages of a job, ($\textit{e.g.}$
shuffle flows in MapReduce). Fig. \ref{fig:MapReduce} shows the shuffle
process in MapReduce, where flows in one shuffle phase are termed
a coflow. The MapReduce shuffle phase will not complete until all
parallel flows have finished transmission. Obviously, the slowest
flow in a coflow critically affects the completion time of reducer
tasks. Hence, coflow completion time (CCT) is the completion time
of the slowest flow within a coflow and minimizing CCT will speed
up the completion of the corresponding job \cite{literature5}.

\begin{figure}[tbh]
\centering \includegraphics[width=0.25\textwidth,height=4.3cm]{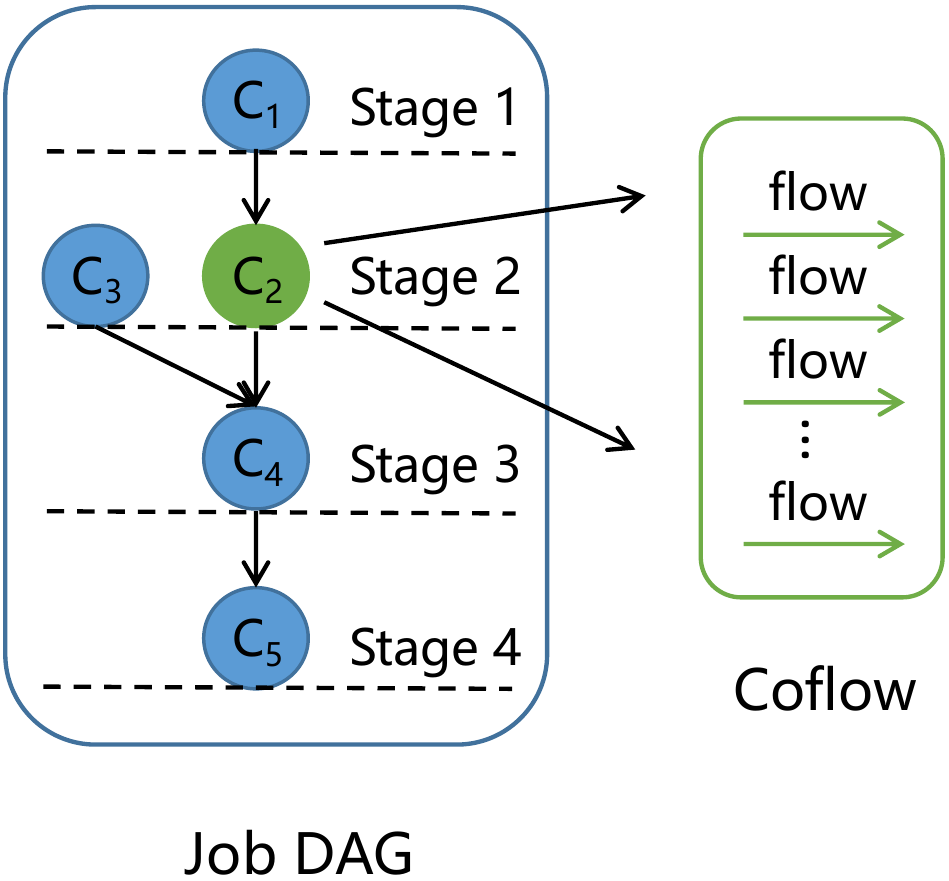}
\caption{Multi-stage job structure}
\label{fig:Multi-stage}
\end{figure}

To improve the communication performance of data-parallel computing
jobs, many scheduling methods \cite{literature6,literature7,literature8,literature9,literature10,literature11}
have been proposed to minimize CCT. These researches have greatly
accelerated the execution of a single-stage job containing only one
coflow, where minimizing CCT implies minimizing job completion time
(JCT). However, in the data-parallel computing framework, multi-stage
jobs are common. The structure of a multi-stage job is shown in Fig.
\ref{fig:Multi-stage}, where the nodes represent the coflows and
the directed edges represent the dependencies between them. Each multi-stage
job contains multiple coflows with dependencies, and each coflow is
composed of multiple parallel flows. There are two types of dependencies:
$\textit{starts-after}$ dependency and $\textit{finishes-before}$
dependency \cite{literature12}, where $\textit{starts-after}$ indicates
that $C_{2}$ cannot start until $C_{1}$ has finished and $\textit{finishes-before}$
indicates that $C_{2}$ can coexist with $C_{1}$ but it cannot finish
until $C_{1}$ has finished.

In the multi-stage job scenarios, minimizing the average CCT does
not necessarily lead to minimizing jobs completion time because the
dependencies in a job should be considered. To tackle this problem,
several heuristics \cite{literature12} and approximation solutions
\cite{literature13,literature15} have recently been proposed to minimize
JCT. However, they simplified the problem with some relaxation and
ignored the workload characteristics. In recent years, with the development
of artificial intelligence, machine learning methods have been increasingly
applied to task scheduling problems in cloud data centers. Reinforcement
learning (RL), as a current research hotspot in the field of machine
learning, is used by many researchers to solve task scheduling problems
in complex data center environments. \cite{literature16,literature17,literature18}.
RL can learn directly from the actual working environment without
relying on inaccurate empirical summaries. Compared with traditional
reinforcement learning, Deep Reinforcement Learning (DRL) \cite{literature19}
has powerful representation capabilities of neural networks and decision-making
capabilities in the field of control.

DeepWeave \cite{literature14} is the first to use the RL framework
to generate efficient coflow scheduling strategies in job DAGs automatically.
In DeepWeave, the node embeddings obtained after GNN processing is
directly flattened into a high-dimensional embedding vector containing
all the information as the input of the policy network. However, this
intuitive approach cannot scale to the job DAG of arbitrary size and
shape, particularly because neural networks usually require fixed-sized
vectors as inputs and processing a high-dimensional feature vector
would require a large policy network that is difficult to train \cite{literature26}.
In addition, DeepWeave considers the scheduling problem of a single
job DAG while ignoring the online situation where multiple multi-stage
jobs coexist, i.e., there is the newly coming job DAG in addition
to the existing ones that have not finished their transmission in
the network.

It has been shown that even if all coflows arrive simultaneously and
their sizes are known in advance, the coflow scheduling problem of
single-stage jobs is NP-hard \cite{literature6}. Therefore, the coflow
scheduling problem of online multi-stage jobs is more challenging.
In this paper, we propose an Attention-based Reinforcement Learning
Model to generate coflow scheduling policies for multi-stage jobs,
which can process DAGs of arbitrary sizes and shape while ensuring
the accuracy of the scheduling strategy. Considering that jobs from
different applications have different importance, jobs with larger
weights should be given priority. In this paper, we take average/total
weighted JCT as the main optimization goal.

Specifically, we employ a directed acyclic graph neural network (DAGNN)
\cite{literature21} to encode DAG information in a set of embedding
vectors and incorporate the partial order relationship in the DAG
into the neural network. However, DAGNN uses an iterative message
passing scheme which is computationally expensive. Consequently, we
propose a Pipeline Mechanism to speed up the feature extraction process,
namely Pipelined-DAGNN. After Pipelined-DAGNN processing, we feed
the embedding sequence of schedulable coflows to the self-attention
layer for information interaction, where a schedulable coflow is one
that has no predecessor dependencies or all its preceding flows have
been scheduled. The output self-attentive sequence serves as the input
of the policy network to generate a priority scheduling list. The
self-attention mechanism is the key for ensuring the accuracy of the
scheduling policy. Based on the list, flows in each coflow is further
scheduled in a fine-grained fashion. In our multi-stage job scheduling
problem, information about the future jobs is not known. Therefore,
our scheduling algorithm is online, which consists of coflow ordering
and scheduling two parts. See Section \ref{sec:algorithm} for details.

The contributions of this paper can be summarized as follows:
\begin{itemize}
\item We propose a novel Pipelined-DAGNN that can effectively speed up the
feature extraction process of DAGNN \cite{literature21} and encode
the job information in a set of embedding vectors.
\item We present a novel scalable deep RL model that feeds an embedding
sequence of vectors to the policy network instead of creating a high-dimensional
flat embedding vector as DeepWeave \cite{literature14}, which significantly
reduces the size of the policy network and hence can process DAGs
of arbitrary sizes and shapes.
\item We propose a novel scheme of incorporating the self-attention mechanism
into the deep RL model to capture the interaction between schedulable
coflows encoded as the embedding sequence, and thus remedy the accuracy
loss in our output policy without packing individual embedding vectors
into a long flat vector.
\item We evaluate the model's performance using real-life traces and generated
traces. Simulation results show our model outperforms state-of-the-art
solutions in terms of the total weighted JCT.
\end{itemize}
The remainder of the paper is organized as follows: Section \ref{sec:rw}
discusses the related work in recent years. Section \ref{sec:sys-all}
mathematically formulates the problem we study. Section \ref{sec:framework}
and Section \ref{sec:algorithm} describe the proposed model and online
scheduling algorithm, respectively. Section \ref{sec:experiment and evaluation}
presents the experimental results of our algorithm on real-world data
collected from Facebook and performance comparison with the state-of-the-art
work. Finally, Section \ref{sec:conclusion} concludes the paper.

\section{Related Work}

\label{sec:rw} Coflow abstraction can better capture application-level
semantics, thereby improving the communication performance of distributed
data-parallel jobs. Existing work mostly focuses on scheduling coflows
in single-stage jobs and minimizing coflow completion time (CCT) as
the optimization goal. However, only a few works consider multi-stage
scenarios, which concentrate on minimizing job completion time (JCT).
Next, we introduce related work from these two aspects.

$\textbf{Single-Stage Scheduling :}$ For the single-stage job scheduling
problem, many heuristic algorithms have been proposed to minimize
CCT, including heuristic algorithms \cite{literature4,literature6,literature7,literature8}.
Orchestra \cite{literature4} is perhaps the first work to mention
the concept of coflow and shows that even a simple FIFO algorithm
can significantly improve coflow performance. Coflow abstraction is
later formally defined in \cite{literature5}. Varys \cite{literature6}
proposes the smallest-effective-bottleneck-first (SEBF) and minimum-allocation-for-desired-duration
(MADD) heuristic algorithms to greedily schedule coflows according
to the bottleneck completion time of coflow to minimize CCT and meet
the deadline of coflow at the same time. Barrat \cite{literature7}
and Stream \cite{literature8} both aim at decentralized coflow scheduling,
and Barrat \cite{literature7} exploits multiplexing to prevent head-of-line
blocking to small coflows. There is also some theoretical work \cite{literature9,literature10,literature11}
aimed at minimizing the approximate algorithm of the total weighted
CCT. \cite{literature9} provided the first deterministic algorithm
with a constant approximation ratio of $\frac{67}{3}$ to minimize
the total weighted CCT. \cite{literature10} presented a 12-approximation
ratio algorithm by reducing to concurrent open shop problem, and \cite{literature11}
further proposed a 5-approximation ratio algorithm by relaxing the
problem to linear programming (LP). The work as mentioned above abstracts
the network topology as a big non-blocking switch without considering
the constraints of network resources. Now, several research works
have considered network resource constraints and combined routing
and resource scheduling to minimize CCT \cite{literature36,literature37}.
Rapier \cite{literature36} is the first algorithm that considers
routing and scheduling jointly to minimize CCT. However, it is a heuristic
solution and cannot provide theoretical performance guarantees. Next,
Tan et al. \cite{literature37} proposed a rounding-based randomization
algorithm for single coflow and online algorithms to provide performance
guarantees for multiple coflow routing and scheduling. Recently, a
distributed bottleneck aware coflow scheduling algorithm \cite{literature38}
was proposed to minimize CCT by reducing network bottlenecks and improving
link capacity utilization. However, these algorithms focus on minimizing
CCT without considering the dependencies among coflows of multi-stage
jobs.

$\textbf{Multi-Stage Scheduling:}$ Although the above methods have
been proven to be effective for single-stage job scheduling problems,
they may not be suitable for solving coflow scheduling problems in
multi-stage jobs. In a multi-stage job scenario, due to the dependency
between coflows, minimizing CCT may not minimize JCT. To the best
of our knowledge, only a few works \cite{literature12,literature13,literature14,literature39}
considered the coflow scheduling problem in the case of multi-stage
jobs. Aalo \cite{literature12} is the first effective heuristic algorithm
with the objective of minimizing JCT, which discusses coflow scheduling
in multi-stage jobs, and proposes to prioritize coflows according
to dependency orders to schedule coflow. However, it only takes a
short section to discuss the heuristic algorithm without neither formal
formulation and analysis. Tian et al. \cite{literature13} proposed
a deterministic algorithm with an $\textit{M}$-approximation ratio
by relaxing the multi-stage job scheduling problem to LP, where $\textit{M}$
represents the number of machines. Unfortunately, it cannot be used
in online cases as all job statistics are required to be known in
advance to solve the relevant LP. DeepWeave \cite{literature14} is
the first to propose a reinforcement learning model to solve the coflow
scheduling problem, which can adaptively generate an efficient coflow
scheduling policy without human expertise. However, this approach
cannot scale to job DAGs of arbitrary sizes and shapes, and ignores
the online situation where multiple multi-stage jobs coexist in the
network. Similarly, \cite{literature39} is the first to study how
to schedule multi-stage jobs under taking network resource constraints
into account so as to minimize the total weighted JCT.

In this paper, we ignore network resource constraints and only consider
the problem of minimizing the total weighted JCT of multi-stage jobs
in a large non-blocking switch.

\section{Formulation}

\label{sec:sys-all} In this section, we first introduce the model
of datacenter networks and multi-stage jobs, and then formulate the
coflow scheduling problem in online multi-stage jobs. The notations
to be used are listed in Tables \ref{tab:notation1} and \ref{tab:notation2}.

\begin{table}[h]
{\small{}{}{}\vspace*{-0.8\baselineskip}
 \caption{Notations of constants.}
\label{tab:notation1} }%
\begin{tabular}{lp{0.8\linewidth}}
\toprule 
{\small{}{}{}Symbol} & {\small{}{}{}Definition}\tabularnewline
\midrule 
{\small{}{}{}$N$} & {\small{}{}{}The number of Jobs}\tabularnewline
{\small{}{}{}$K_{n}$} & {\small{}{}{}The number of coflows in job $j_{n}$}\tabularnewline
{\small{}{}{}$P$} & {\small{}{}{}The size of network}\tabularnewline
{\small{}{}{}$M_{n,k}$} & {\small{}{}{}The number of flows in coflow $c_{n,k}$}\tabularnewline
{\small{}{}{}$X_{n,k,k^{\prime}}$} & {\small{}{}{}The binary constant indicates whether there is a dependency:
coflow $c_{n,k}$ starts after coflow $c_{n,k^{\prime}}$}\tabularnewline
{\small{}{}{}$a_{n}$} & {\small{}{}{}The arrival time of job $j_{n}$}\tabularnewline
{\small{}{}{}$w_{n}$} & {\small{}{}{}The weight of job $j_{n}$}\tabularnewline
{\small{}{}{}$b_{n,k,m}$} & {\small{}{}{}The total bytes of flow $f_{n,k,m}$}\tabularnewline
\bottomrule
\end{tabular}{\small{}{}{}\vspace*{-1\baselineskip}
}
\end{table}

\begin{table}[h]
{\small{}{}{}\vspace*{-0.2\baselineskip}
 \caption{Notations of variables.}
\label{tab:notation2} }%
\begin{tabular}{lp{0.8\linewidth}}
\toprule 
{\small{}{}{}Symbol} & {\small{}{}{}Definition}\tabularnewline
\midrule 
{\small{}{}{}$J_{n}$} & {\small{}{}{}The completion time of job $j_{n}$}\tabularnewline
{\small{}{}{}$C_{n,k}$} & {\small{}{}{}The completion time of coflow $c_{n,k}$}\tabularnewline
{\small{}{}{}$F_{n,k,m}$} & {\small{}{}{}The completion time of flow $f_{n,k,m}$}\tabularnewline
{\small{}{}{}$\overline{F}_{n,k,m}$} & {\small{}{}{}The starting transmission time of flow $f_{n,k,m}$}\tabularnewline
{\small{}{}{}$r_{n}$} & {\small{}{}{}The release time of job $j_{n}$}\tabularnewline
\bottomrule
\end{tabular}{\small{}{}{}\vspace*{-1\baselineskip}
}
\end{table}

\subsection{Model}

\begin{figure}[h]
\includegraphics[width=0.45\textwidth,height=4cm]{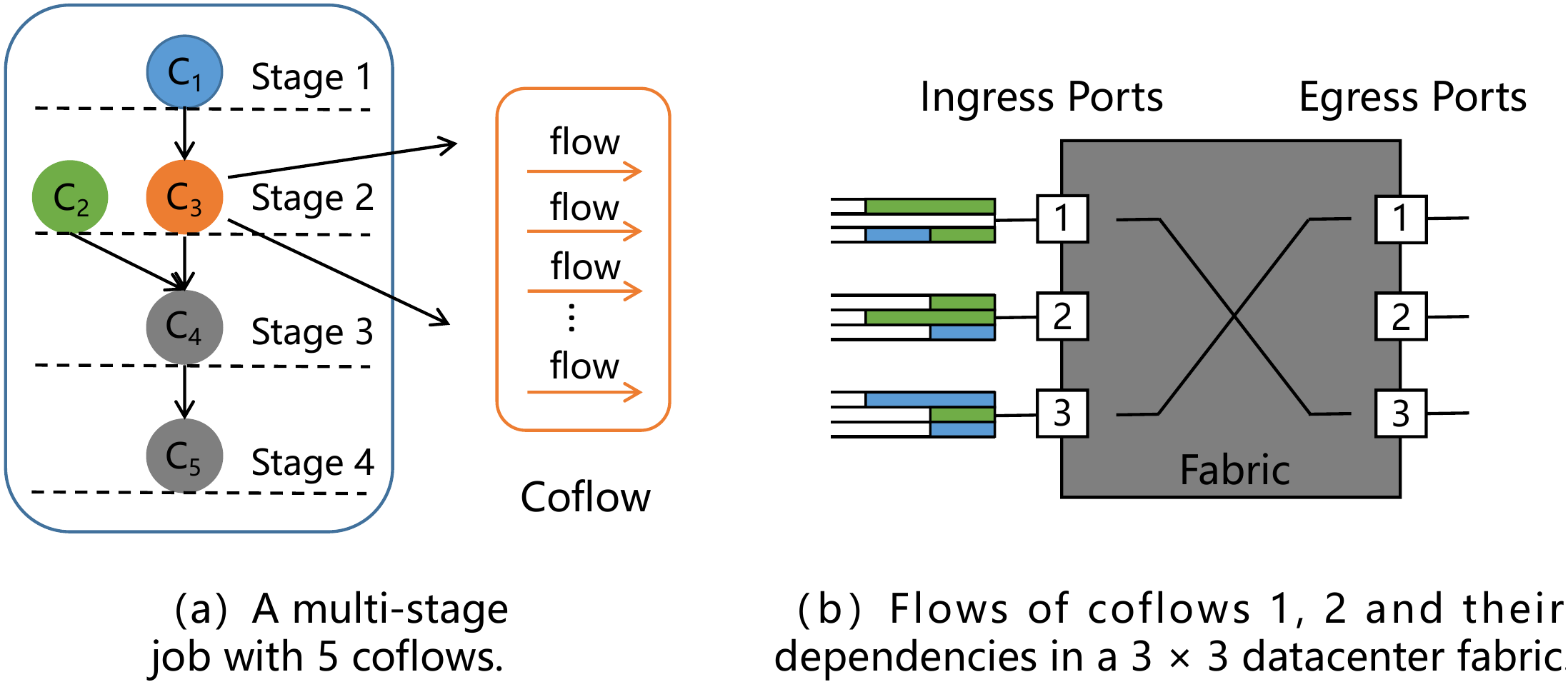}
\caption{Coflow scheduling over a 3 $\times$ 3 data center fabric}
\label{fig:Big Switch}
\end{figure}

$\textit{Network Model: }$ We extract the network topology as a $\textit{P}$
$\times$ $\textit{P}$ non-blocking switch fabric, with $\textit{P}$
ingress ports connected to $\textit{P}$ source servers and $\textit{P}$
egress ports connected to $\textit{P}$ destination servers. Without
loss of generality, we assume that each ingress/egress port has a
capacity of 1Gbps, equivalently 128MBps. In the online simulation,
we choose the time unit as 1/128 second accordingly so that each port
has a capacity of 1MB per time unit. Fig. \ref{fig:Big Switch} shows
the scheduling of coflows through a 3 $\times$ 3 datacenter fabric.
Under this model, each ingress port has flows from one or more coflows
to various egress ports, port-sharing is allowed, and it is assumed
that congestion occurs only at the ingress and egress ports. Coflows
1 and 2 can share the network resources at the same time because they
are independent. Once all their flows are transmitted, coflow 3 will
be ready to be transmitted. According to the $\textit{starts-after}$
dependency, only when all flows in coflow 1 and coflow 2 are completed,
coflow 3 can be released.

$\textit{Job Model: }$ Considering $N$ jobs in the network, denoted
by $j_{1},j_{2},\ldots,j_{N}$, arriving at $a_{1},a_{2},\ldots,a_{N}$
respectively, and the arrival time of each job obeys Poisson distribution.
In the context of our multi-stage jobs, information about future jobs
is not known. The information of each job $j_{n}$ containing $K_{n}$
coflows, denoted by $c_{n,1},c_{n,2},\ldots,c_{n,K_{n}}$, is given
at arrival. Each coflow $c_{n,k}$ contains $M_{n,k}$ parallel flows,
denoted by $f_{n,k,1},f_{n,k,2},\ldots,f_{n,k,M_{n,k}}$, and the
total bytes of flow $f_{n,k,m}$ is $b_{n,k,m}$.

\subsection{Problem formulation}

In the scenario of multi-stage jobs, we want to optimize the overall
performance of weighted jobs by minimizing the total weighted JCT.

\begin{equation}
\begin{aligned}\label{eq:weightedJCT}\min\sum_{n=1}^{N}w_{n}J_{n}\end{aligned}
\end{equation}
where $J_{n}$ denotes the completion time of job $j_{n}$, the weights
capture different priorities for different jobs and important jobs
are prioritized by assigning higher weights. Therefore, minimizing
the average JCT is a special case of our goal when all weights are
equal.

The completion time of a multi-stage job depends on the completion
time of the latest coflow that composes it. So 
\begin{equation}
\begin{aligned}\label{eq:JCT}J_{n}=\max_{k\in\left\{ 1,2,\ldots,K_{n}\right\} }C_{n,k},\forall n\in\{1,2,\ldots,N\}\end{aligned}
\end{equation}
where $C_{n,k}$ represents the completion time of coflow $c_{n,k}$.

Similarly, the completion time of a coflow depends on the completion
time of the latest flow that composes it. Hence, 
\begin{equation}
\begin{aligned}\label{eq:CCT}C_{n,k}=\max_{m\in\left\{ 1,2,\ldots,M_{n,k}\right\} }F_{n,k,m},\forall n\in\{1,2,\ldots,N\},\\
k\in\left\{ 1,2,\ldots,K_{n}\right\} 
\end{aligned}
\end{equation}
where $F_{n,k,m}$ represents the completion time of flow $f_{n,k,m}.$

$\textbf{Start time constraints:}$ Flows in a job can start its transmission
only after the job is released. So 
\begin{equation}
\begin{aligned}\label{eq:Starttime}\overline{F}_{n,k,m}\geq r_{n},\\
\forall n\in\{1,2,\ldots,N\},k\in\left\{ 1,2,\ldots,K_{n}\right\} ,\\
m\in\left\{ 1,2,\ldots,M_{n,k}\right\} 
\end{aligned}
\end{equation}
where $\overline{F}_{n,k,m}$ denotes the starting transmission time
of flow $f_{n,k,m}$ and $r_{n}$ denotes the release time of job
$j_{n}$. Obviously, $a_{n}\leq r_{n}$, $a_{n}$ denotes the arrival
time of job $j_{n}$.

$\textbf{Dependency constraints:}$ In this paper, we only consider
scheduling multi-stage jobs with $\textit{starts-after}$ dependencies.
The flows in a coflow cannot start their transmission until the coflows
on which they depend have completed their transmission. Consequently,
the dependency constraint can be expressed as follows 
\begin{equation}
\begin{aligned}\label{eq:Dependency}\overline{F}_{n,k,m}\geq X_{n,k,k^{\prime}}C_{n,k^{\prime}},\\
\forall n\in\{1,2,\ldots,N\},k,k^{\prime}\in\left\{ 1,2,\ldots,K_{n}\right\} ,\\
m\in\left\{ 1,2,\ldots,M_{n,k}\right\} 
\end{aligned}
\end{equation}
where $X_{n,k,k^{\prime}}$ is a binary indicator which is equal to
1 if there is a dependency that coflow $c_{n,k}$ starts after coflow
$c_{n,k^{\prime}}$ or 0 if otherwise \cite{literature14}.

\section{The Framework}

\label{sec:framework} In this section, we introduce the framework
of the proposed model. First, we overview the architecture of the
model. Second, we introduce the processing mechanism of DAGNN, and
present our Pipelined-DAGNN to speed up the feature extraction process.
Then, we give the Self-Attention Mechanism which is the key for ensuring
the performance of the coflow scheduling strategy. Finally, we describe
the training framework of the DRL model.

\begin{figure*}[tbh]
\centering \includegraphics[width=0.6\textwidth,height=5cm]{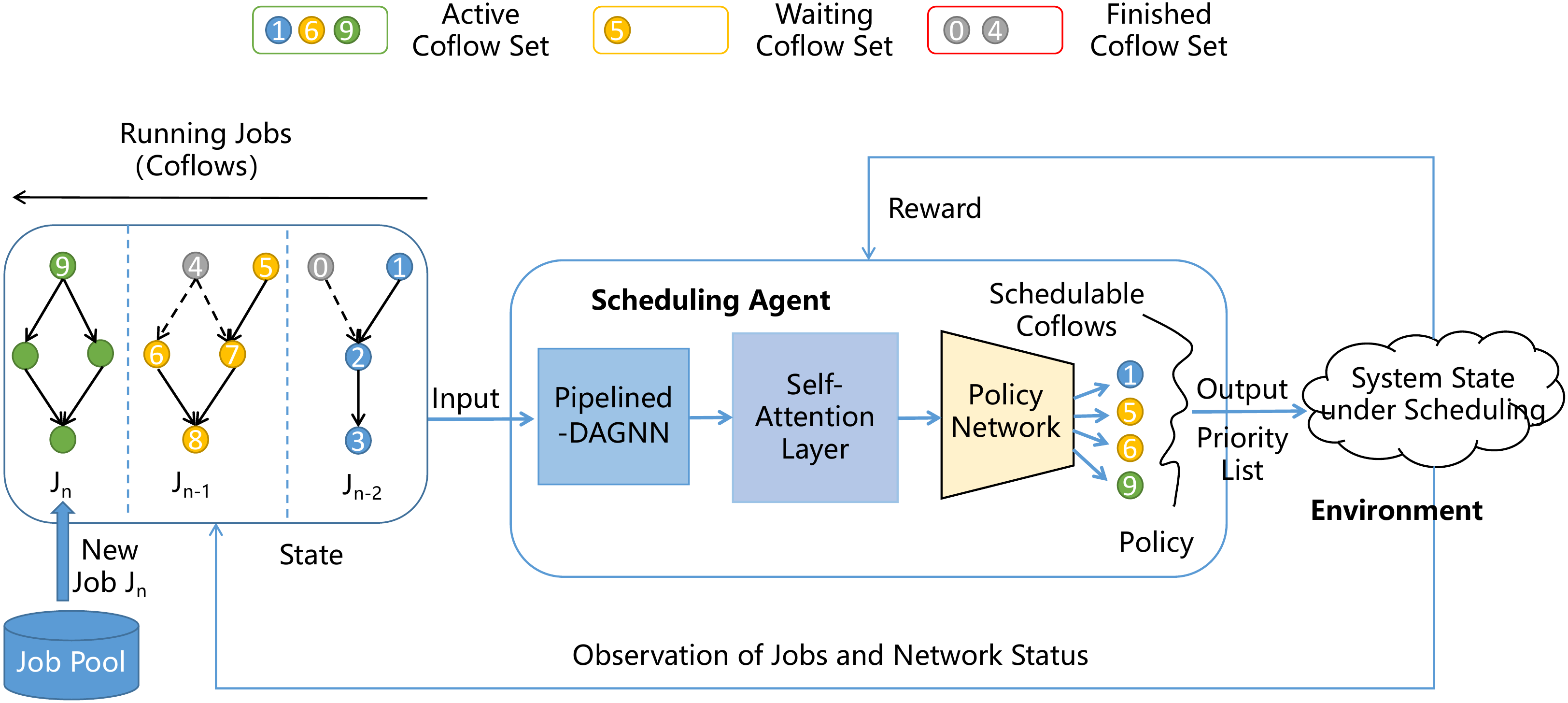}
\caption{The DRL framework with Self-Attention Mechanism}
\label{fig:DRL Agent}
\end{figure*}

\subsection{Overview of the Framework}

The proposed model is designed based on the framework of DRL and can
automatically generate efficient coflow scheduling policies for the
network. Fig. \ref{fig:DRL Agent} is the complete learning framework
of the proposed model, where the agent plays the role of the coflow
scheduler, which is composed of three modules, Pipelined-DAGNN, Self-Attention
Layer and Policy Network. (See Section \ref{sec:Pipelined-DAGNN Processing},
Section \ref{sec:Self-Attention Processing} and Section \ref{sec:Policy Network Processing}
for details).

When the model is running, the agent collects the job DAGs information
and converts the information to a feature map as the input of Pipelined-DAGNN,
which encodes the features of the job DAGs in a set of embeddings
and passes the embedding sequence composed of schedulable coflows
to a Self-Attention Layer. Finally, the generated self-attentive sequence
serves as the input to the Policy Network and outputs a priority list
for scheduling coflows and the flows in coflows. The active queue
stores coflows that have been released but not yet completed; the
coflows in the waiting queue need to wait for higher priority coflows
to release the required port resources. All currently schedulable
coflows are placed in the active queue and waiting queue, respectively.
With the release of resources, the coflows that have finished transmission
are placed in finished queue, the coflows in the waiting queue will
enter the active queue , but the coflows in the active queue cannot
enter the waiting queue because preemption is not allowed.

\subsection{Neural Network Implementation}

\begin{figure*}[tbh]
\centering \includegraphics[width=0.55\textwidth,height=3.5cm]{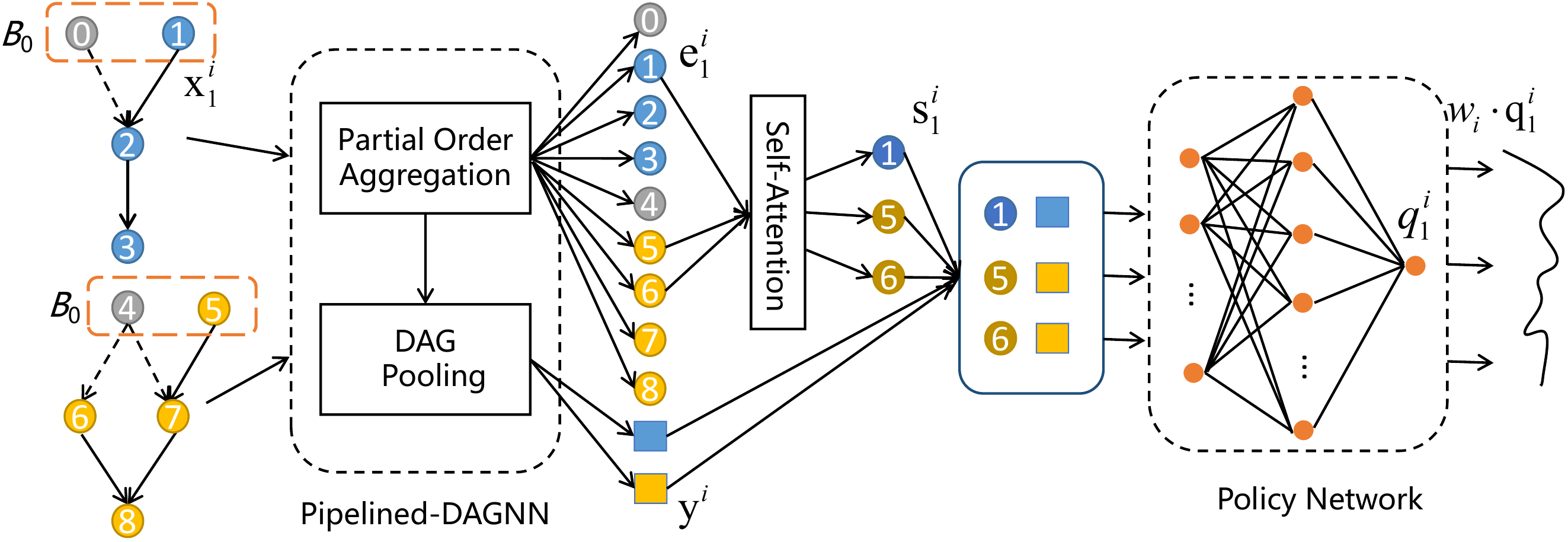}
\caption{Neural Network Implementation}
\label{fig:Neural Network}
\end{figure*}

As shown in Fig. \ref{fig:Neural Network}, the neural network implementation
in our model includes three stages: Pipelined-DAGNN processing, self-attention
processing, and policy network processing. Next, we describe the three
stages in detail.

\subsubsection{Pipelined-DAGNN Processing}

\label{sec:Pipelined-DAGNN Processing} In multi-stage jobs, each
job consists of multiple coflows and is represented by a Directed
Acyclic Graph (DAG). Compared with the most common GNN architecture
that aggregates information from neighbouring nodes at the previous
layer based on message passing, DAGNN is driven by the partial order
induced by the DAG and aggregates neighbourhood information in the
current layer \cite{literature21}. Note that DAGNN exploits an inductive
bias that relies on the assumption that the input graph is a DAG.
To facilitate subsequent understanding, we first briefly describe
the DAGNN framework:

\begin{equation}
\begin{aligned}\label{eq:DAGNN1}H_{v}^{\ell}=F^{\ell}\left(G^{\ell}\left(\left\{ H_{u}^{\ell}\mid u\in\mathcal{P}(v)\right\} ,H_{v}^{\ell-1}\right),H_{v}^{\ell-1}\right),\\
\ell=1,\ldots,L
\end{aligned}
\end{equation}

\begin{equation}
H_{\mathcal{G}}=R\left(\left\{ H_{v}^{\ell},\ell=0,1,\ldots,L,v\in\mathcal{T}\right\} \right)\label{eq:DAGNN2}
\end{equation}

Eq. \eqref{eq:DAGNN1} computes representations $H_{v}^{\ell}$ for
all nodes $v$ in a graph $\mathcal{G}$ in every layer $\ell$ and
Eq. \eqref{eq:DAGNN2} readouts the whole graph representation $H_{\mathcal{G}}$,
where $\mathcal{P}(v)$ denotes the set of direct predecessors of
$v,\mathcal{T}$ denotes the set of nodes without successors, and
$G^{\ell}(\cdot),F^{\ell}(\cdot)$, and $R(\cdot)$ are parameterized
neural networks \cite{literature21}.

\begin{figure}[h]
\centering \includegraphics[width=0.16\textwidth,height=3cm]{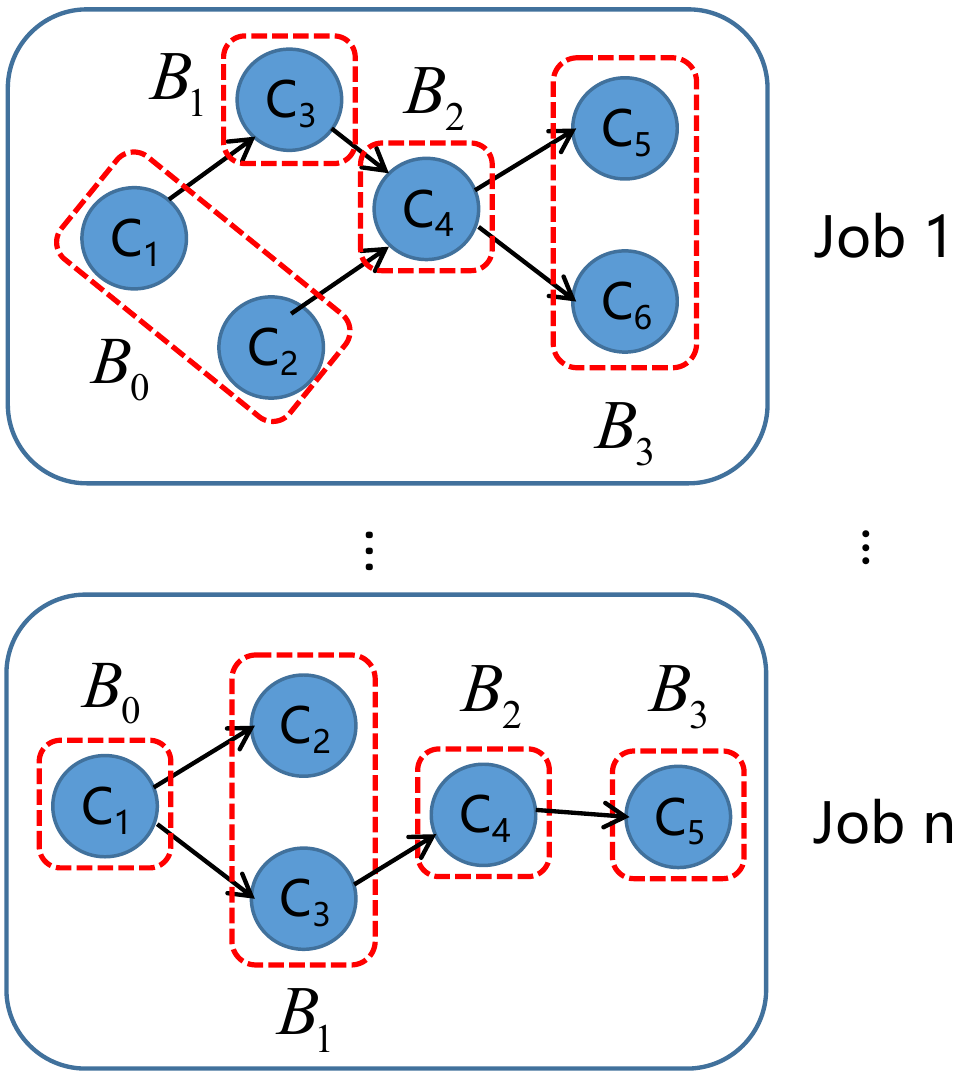}
\caption{Topological Batching for Multiple Job DAGs}
\label{fig:Partition}
\end{figure}

It can be seen from the above formula, DAGNN uses only direct predecessors
for aggregation and the pooling on only nodes without successors.
In addition, as shown in Fig. \ref{fig:Partition}, DAGNN considers
$\textit{topological batching}$, which divides all nodes in job DAG
into sequential batches $\textit{B}_{i}$ that nodes without dependency
may be grouped together and processed concurrently if their predecessors
have all been processed \cite{literature21}. Obviously, sequential
batch $\textit{B}_{i}$ of the same $i$ in different DAGs can be
combined and processed simultaneously, which achieves better parallel
concurrency. It can be seen from Fig. \ref{fig:Neural Network} that
nodes 0, 1, 4, and 5 come from different jobs in the same sequential
batch $\textit{B}_{0}$, which can be processed concurrently by DAGNN.

Inspired by \cite{literature21}, we employ DAGNN to encode the job
DAGs information in a set of embeddings, which contain per-node embedding
$\mathbf{e}_{v}^{i}$ and per-job embedding $\mathbf{y}^{i}$. The
former captures information about the node and its predecessor nodes
and the latter aggregates information across the set of nodes without
successors $\mathcal{T}$. Given the feature vectors $\mathbf{x}_{v}^{i}$
of node $v$ in job DAG $G_{i}$, we build a per-node embedding $\left(G_{i},\mathbf{x}_{v}^{i}\right)\longmapsto\mathbf{e}_{v}^{i}$,
and compute a per-job embedding for each DAG $G_{i}$, $\left\{ \left(\mathbf{x}_{v}^{i},\mathbf{e}_{v}^{i}\right),v\in\mathcal{T}_{i}\right\} \longmapsto\mathbf{y}^{i}.$
Table \ref{tab:notation3} defines our notation. However, DAGNN is
still computationally expensive since it uses an iterative message
passing scheme. Hence, we propose a novel Pipelined-DAGNN that can
effectively speed up the feature extraction process of DAGNN.

\begin{table}[!t]
{\small{}{}{}\vspace*{-0.2\baselineskip}
 \caption{Notations.}
\label{tab:notation3} }%
\begin{tabular}{lp{0.8\linewidth}}
\toprule 
{\small{}{}{}Symbol} & {\small{}{}{}Definition}\tabularnewline
\midrule 
{\small{}{}{}$i$} & {\small{}{}{}$i$-th job}\tabularnewline
{\small{}{}{}$v$} & {\small{}{}{}DAG node (coflow)}\tabularnewline
{\small{}{}{}$G_{i}$} & {\small{}{}{}job $i$'s DAG}\tabularnewline
{\small{}{}{}$\mathbf{x}_{v}^{i}$} & {\small{}{}{}per-node feature vector}\tabularnewline
{\small{}{}{}$\mathbf{e}_{v}^{i}$} & {\small{}{}{}per-node embedding}\tabularnewline
{\small{}{}{}$\mathbf{y}^{i}$} & {\small{}{}{}per-job embedding}\tabularnewline
{\small{}{}{}$\mathbf{s}_{v}^{i}$} & {\small{}{}{}per-node self-attentive embedding}\tabularnewline
{\small{}{}{}$\textit{q}_{v}^{i}$} & {\small{}{}{}per-node priority value}\tabularnewline
{\small{}{}{}$w_{i}$} & {\small{}{}{}the weight of job $j_{i}$}\tabularnewline
\bottomrule
\end{tabular}{\small{}{}{}\vspace*{-1\baselineskip}
}
\end{table}

\begin{figure}[h]
\includegraphics[width=0.45\textwidth,height=3cm]{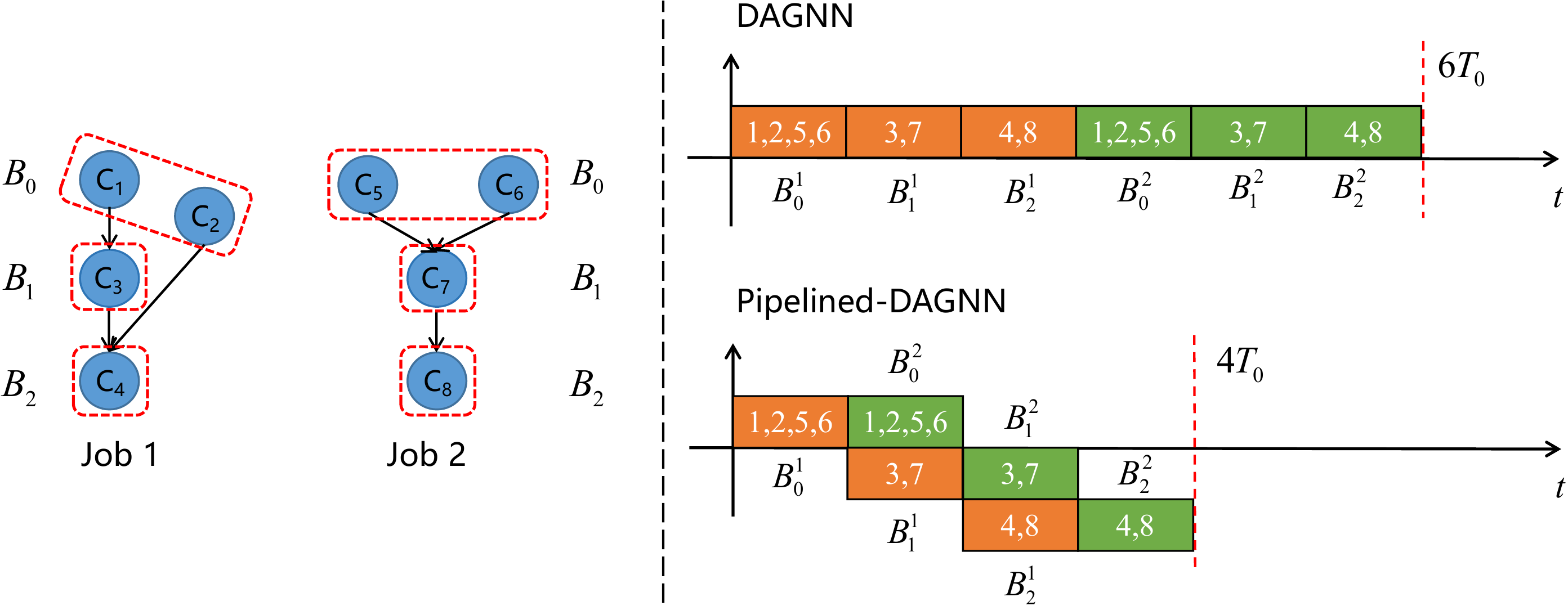} \caption{Pipeline Mechanism Acceleration}
\label{fig:Pipeline}
\end{figure}

To demonstrate this, let us consider a simple example in Fig. \ref{fig:Pipeline}.
Using topological batching to update the two job DAG in the figure
at the same time, and assuming that the average time required for
$\textit{B}_{j}$ to be updated once is $\textit{T}_{0}$. $\textit{L}$
= 2, is the number of iterations. $\textit{N}$ = 3, is the maximum
number of sequential batches in job DAGs. Because Pipelined-DAGNN
use the current-layer, rather than the past-layer, information to
compute the current-layer representation of $v$ and aggregates from
the direct-predecessor set $\mathcal{P}(v)$ only, rather than the
entire neighborhood \cite{literature21}. As shown in Fig. \ref{fig:Partition},
$\textit{B}_{j}^{1}$ and $\textit{B}_{j}^{2}$ respectively represent
the first iteration update and the second iteration update of the
sequential batch $\textit{B}_{j}$. The time required for sequential
batch $\textit{B}_{j}$ to iteratively update twice is 6$\textit{T}_{0}$,
and considering the pipeline mechanism, when $\textit{B}_{0}^{1}$
ends, $\textit{B}_{0}^{2}$ and $\textit{B}_{1}^{1}$ can be performed
simultaneously, the time spent is 4$\textit{T}_{0}$. It can be inferred
from the figure that the time spent without considering the pipeline
mechanism is $\textit{N}\times\textit{L}\times\textit{T}_{0}$. On
the contrary, the time spent is $(\textit{N}-1)\times\textit{T}_{0}+\textit{L}\times\textit{T}_{0}$.
The acceleration ratio is $\frac{\textit{L}\times\textit{N}}{\textit{L}+\textit{N}-1}$.

\subsubsection{Self-Attention Processing}

\label{sec:Self-Attention Processing} After Pipelined-DAGNN processing,
the embedding of each node and each job in DAGs can be obtained. However,
for job DAGs, there is a dependency relationship between nodes, and
only those that have no predecessor dependencies or all predecessor
dependencies that have finished scheduling can be scheduled. As shown
in Fig. \ref{fig:Neural Network}, node 1 has no predecessor dependency,
and the predecessor node 0 of node 4 has completed scheduling. Therefore,
the current schedulable nodes are 1, 5, 6. Next, we consider feeding
the embedding sequence of current schedulable nodes to the policy
network to output a priority sequence. In this way, the scale of the
policy network has nothing to do with the number of jobs or nodes,
only the feature dimension of node embedding, so as to reduce the
size of the policy network and achieve scalability. But no matter
from the same job DAG or different job DAGs, there is no directed
edge connection between the current schedulable nodes. Each schedulable
node embedding only contains its own information and information aggregated
from direct-predecessor nodes after Pipelined-DAGNN processing. Therefore,
the priority scores output by the policy network in parallel are independent,
and the generated priority sequence cannot be used as an effective
strategy to guide the agent, resulting in the low accuracy of the
coflow scheduling algorithm. To solve it, we utilize the self-attention
mechanism as an operation to calculate the relevance between different
parts of the embedding sequence.

Intuitively, the attention mechanism is similar to the human attention
distribution process, that is, in the information processing process,
different content is assigned different attention weights. The essence
of Attention can be described as a query $Q$ to a series of key-value
($K$-$V$) pairs of mapping. Self-Attention \cite{literature22}
is a special form of Attention Mechanism. The self-attention model
is actually to make $Q$, $K$ and $V$ equal, and do attention inside
the sequence to find the connections within the sequence. Essentially,
for each input vector, Self-Attention generates a vector that is weighted
and summed on its neighboring vectors, where the weight is determined
by the relationship between inputs. For example, if we enter a sentence,
each word in it must be calculated with all the words in the sentence.
The purpose is to learn the word relevance within the sentence and
capture the internal structure of the sentence.

Feeding the embedding sequence $Z_{embed}$ of currently schedulable
nodes to the self-attention layer is achieved as follows: First, perform
a linear projection on $Z_{embed}$, that is, assign three weight
matrices $W_{Q}$,$W_{K}$,$W_{V}$$\in$$\mathbb{R}^{embed.dim*embed.dim}$
to $Z_{embed}$ to obtain three matrices $Q$,$K$,$V$ representing
queries, keys, and values respectively, with the same dimensions.
Then the similarity is calculated by $\sqrt{d_{k}}$-regulated dot-product
to obtain the weights ($QK^{T}/\sqrt{d_{k}}$), where $d_{k}$ is
the dimension of the key vectors serving as a scaling factor \cite{literature22},
and the softmax function is applied to normalize these weights; Finally,
the weights and corresponding value (matrix) $V$ are multiplied to
obtain the attention. This process can be done via matrix multiplications:
\begin{align*}
 & Q=Linear(Z_{embed})=Z_{embed}W_{Q}\\
 & K=Linear(Z_{embed})=Z_{embed}W_{K}\\
 & V=Linear(Z_{embed})=Z_{embed}W_{V}
\end{align*}

\begin{equation}
Attention(Q,K,V)=\operatorname{soft}\max\left(\frac{QK^{T}}{\sqrt{d_{k}}}\right)V.\label{eq:Self-Attention}
\end{equation}

\subsubsection{Policy Network Processing}

\label{sec:Policy Network Processing} After the self-attention processing
stage, the self-attentive sequence of schedulable node embeddings
and their job level information (per-job embeddings) are sent to the
policy network. For a node $v$ in job DAG $G_{i}$, the policy network
computes a score $q_{v}^{i}\triangleq q\left(\mathbf{s}_{v}^{i},\mathbf{y}^{i}\right)$,
where $q(\cdot)$ is a $score$ $function$ that maps the embedding
(per-node self-attentive embedding and per-job embedding) to a scalar
value. The scalar value is the priority value when the jobs are of
equal weight.

\subsection{DRL Neural Network Training}

\begin{figure}[h]
\centering \includegraphics[width=0.44\textwidth,height=3.8cm]{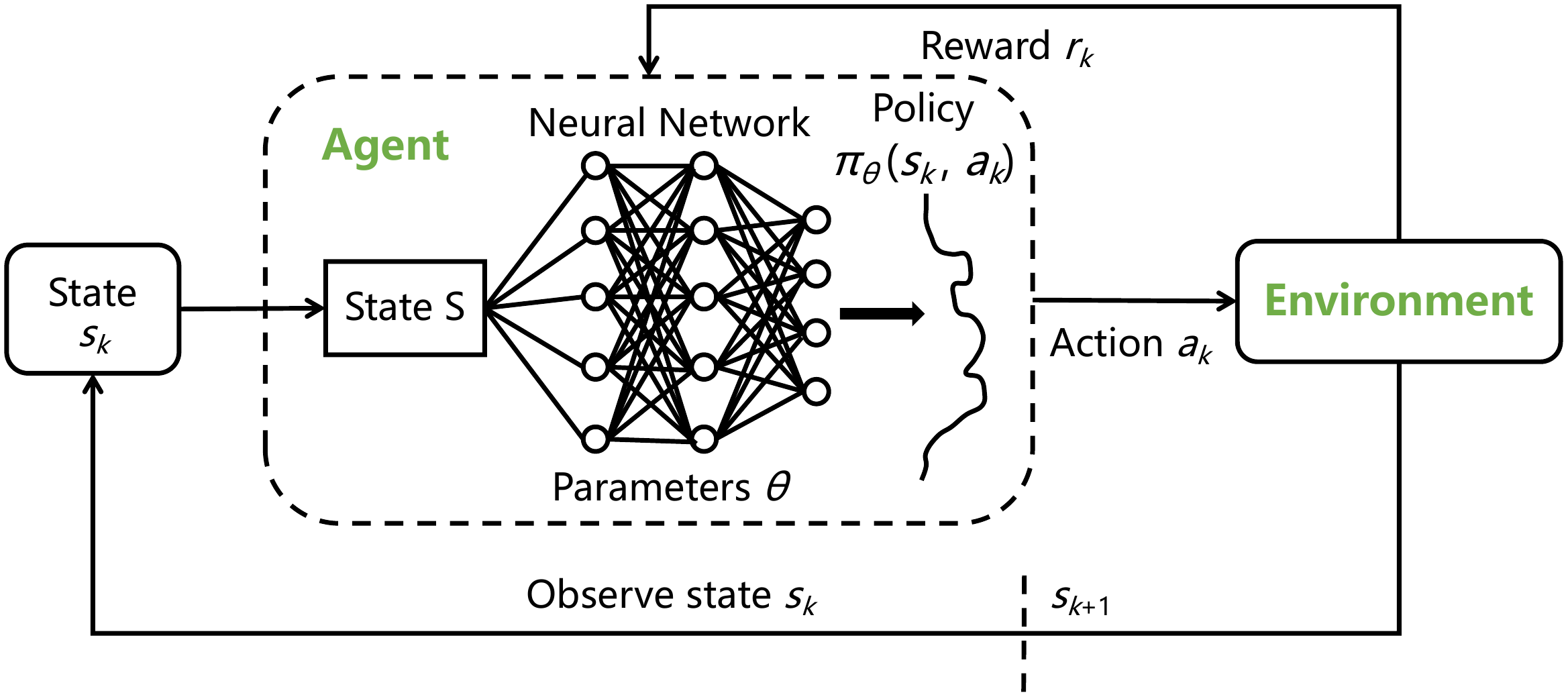}
\caption{The general framework of DRL}
\label{fig:RL}
\end{figure}

Another key task in our model is to train the neural networks in the
framework of DRL comprising of three stages of Pipelined-DAGNN processing,
self-attention processing, and policy network processing, as illustrated
in Fig. 5. These three stages, collectively representing the scheduling
agent in our DRL model, are realized by their respective neural networks
and updated in the end-to-end online training process, with the output
of one stage being the input of another stage.Deep Reinforcement Learning
(DRL) combines the perception ability of deep learning with the decision-making
ability of reinforcement learning and realizes the end-to-end process
from perception to decision-making, which is an artificial intelligence
method closer to human thinking. Consider the general setting in Fig.
\ref{fig:RL}, where a DRL Agent interacts with an Environment. The
DRL agent gets trained through the interaction with the network environment
step by step. At each step $k$, the agent observes a state $s_{k}$
as the input of the parameterized neural network, and the output policy
$\pi_{\theta}(s_{k},a_{k})$ is defined as the probability of taking
action $a_{k}$ in the state $s_{k}$. The agent selects action $a_{k}$
according to the strategy $\pi_{\theta}(s_{k},a_{k})$ to act on the
environment and change the state of the environment. The environment
state is converted to $s_{k+1}$, and the agent receives an instant
reward $r_{k}$ as feedback to evaluate the quality of the action
$a_{k}$.

In our DRL model, the scheduling agent observes the state of the environment,
including the shape of the job DAGs, flows in each coflow and their
occupation of network resources, and outputs the priority list of
the current schedulable nodes. To guide the movement of the agent's
action towards the goal (minimizing the average/total weighted JCT),
the environment will give the agent a reward $r_{k}$ after each action
(ordering). We penalize the agent with a reward $r_{k}=\frac{\sum_{i=1}^{n_{k}}w_{i}}{t_{k+1}-t_{k}}$
after the $k^{th}$ action, where $t_{k}$ is the wall-clock time
of the $k^{th}$ action and $n_{k}$ is the number of jobs that left
in the system during the interval $\left[t_{k},t_{k+1}\right)$, $w_{i}$
is the weight of the corresponding job $i$. RL training proceeds
in episodes. Each episode consists of multiple coflow ordering (action)
and scheduling events. In each episode, the target of the training
is to maximize the cumulated reward $\sum_{k=0}^{T}r_{k}$, where
$T$ denotes the total number of actions performed in this episode.

In our model, we use a policy gradient algorithm \cite{literature23}
for training. In the training process, the policy gradient algorithm
uses the reward of environmental feedback to perform gradient descent
on the neural network parameters to learn. Consider an episode of
length $T$, the agent collects $\left(s_{k},a_{k},r_{k}\right)$
at each step $k.$ The parameters $\theta$ of its policy $\pi_{\theta}\left(s_{k},a_{k}\right)$
is updated as follows

\begin{equation}
\theta\leftarrow\theta+\alpha\sum_{k=1}^{T}\nabla_{\theta}\log\pi_{\theta}\left(s_{k},a_{k}\right)\left(\sum_{k^{\prime}=k}^{T}r_{k^{\prime}}-b_{k}\right)\label{eq:policy gradient}
\end{equation}
where $\alpha$ is the learning rate and $b_{k}$ is a baseline used
to reduce the variance of the policy gradient \cite{literature25}.
The entire training process is shown in Algorithm 4, using the REINFORCE
policy gradient algorithm \cite{literature23} to minimize the parameters
of the neural network and output an effective coflow scheduling strategy.
The key idea of the policy gradient algorithm is to estimate the gradient
using the trajectories of execution with the current policy.

Next, we describe how we implement the algorithm to train our model
in practice. At the beginning of training, the neural network parameters
are initialized randomly, resulting in a poor output policy in the
early stage. With the arrival of jobs, there will be a large number
of queued coflows in almost every training episode. Allowing the agent
to explore beyond a few steps in these early stages wastes training
time. Hence, we can use the method of sampling the episode length
$l$ from an exponential distribution with a small initial mean $l_{mean}$
to terminate the initial episodes early to avoid wasting training
time and hence accelerate the training process. Then, with the multiple-stage
jobs' arrival, the agent follows the Monte Carlo Method to collect
the online experience episodes for training (line 5). Line 12 is the
REINFORCE policy gradient algorithm described in \eqref{eq:policy gradient}.
Line 15 is to increase the average episode length. Finally, we update
the policy parameter $\theta$ on line 16. Our policy network is implemented
using a neural network of two hidden layers with 32 and 16 hidden
units, respectively. We set the learning rate $\alpha$ at $1\times10^{-3}$
and use Adam optimizer for gradient descent. We train our model for
at least 40,000 iterations.

\section{The algorithm}

\label{sec:algorithm} 
\begin{figure}[h]
\centering

\includegraphics[width=0.45\textwidth,height=4.3cm]{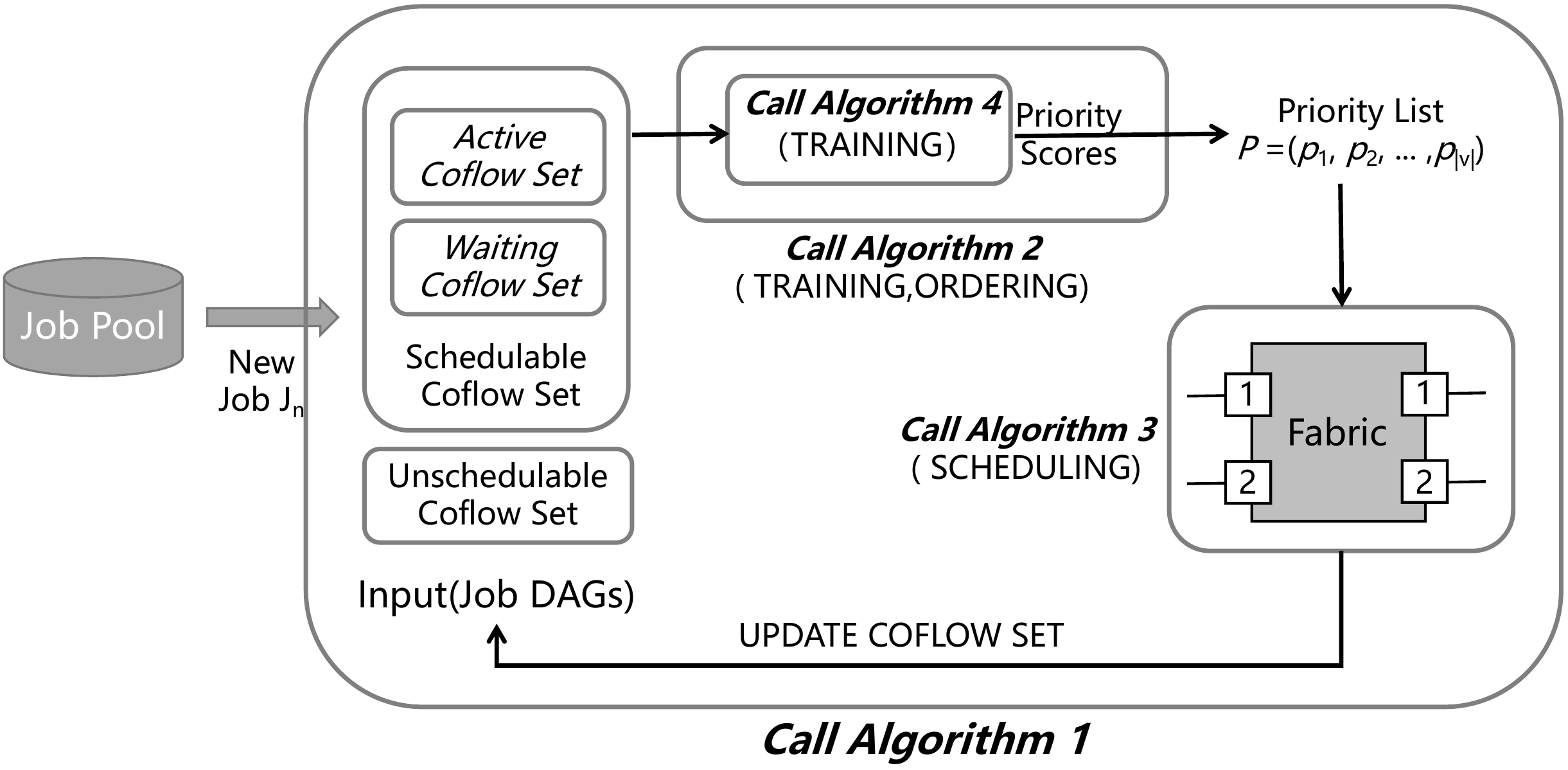}
\caption{Algorithm Design Block Diagram}
\label{fig:four-algorithm}
\end{figure}

In this section, we present an efficient coflow
scheduling algorithm for online multi-stage jobs, as shown in Algorithm
1, which consists of two main components: coflow ordering and scheduling,
respectively implemented by Algorithm 2 and Algorithm 3. Algorithm
2 is responsible for assigning priorities to currently schedulable
coflows based on the DRL framework. Algorithm 3 is responsible for
scheduling flows in active coflows, i.e., the coflows that have been
released but not yet completed among the scheduled coflows, based
on the priority list. The detailed algorithm design block diagram
is shown in Fig. \ref{fig:four-algorithm}.

The following first gives an overall description of the online algorithm,
and then introduces these two components separately.

\subsection{Online Algorithm Design}

Our online scheduling algorithm is shown in Algorithm 1. In the coflow
scheduling problem of online multi-stage jobs, there may be multiple
multi-stage jobs coexisting in the network, thus competing for network
resources. Consequently, we need to determine the priority order of
schedulable coflows. Next, we need to allocate bandwidth resources
to active coflows based on the priority list. In coflow ordering,
a higher value the policy network outputs represent the higher priority,
and Algorithm 2 is called only when a job is released or a coflow
finishes. In coflow scheduling, port-sharing is allowed, and Algorithm
3 is called only when a flow is completed, or a coflow is released.
We first initialize all ports states to idle. 
\begin{algorithm}[H]
\caption{Online scheduling policy for coflows and flows}
\begin{algorithmic}[1] \Require Coflows in the job and job DAGs,
active coflow set $\mathbb{C}$ \Ensure flow rate and flow scheduling
priority \Linecomment INITIALIZE \State Initialize the bandwidth
on each ingress and egress port: $I_{p}=E_{p}=$ port physical bandwidth
\Linecomment ORDERING \State Determining priority list $\leftarrow$
call Algorithm 2. \Linecomment SCHEDULING \State Scheduling active
coflows $\leftarrow$ call Algorithm 3. \If{a flow $F$ completes
transmission or a coflow $C$ is released} \Linecomment UPDATE \State
Update active coflow set $\mathbb{C}$ \Linecomment SCHEDULING \State
Scheduling active coflows $\leftarrow$ call Algorithm 3. \EndIf
\If{a coflow $C$ completes transmission or a job $J$ is released}
\Linecomment UPDATE \State Update active coflow set $\mathbb{C}$
\Linecomment ORDERING \State Determining priority list $\leftarrow$
call Algorithm 2. \EndIf \end{algorithmic} \label{alg:alg1}
\end{algorithm}

\subsection{Determining Priority List}

In this paper, we train the DRL model based on the policy gradient
strategy by Algorithm 4. In the initial stage of training, the network
parameters are random, and the output strategy is poor, resulting
in a large number of job waiting. To avoid wasting training time with
useless exploration in the early stage, we sample the episode length
$\textit{l}$ from an exponential distribution with a small initial
mean $\textit{l}_{\text{mean}}$ \cite{literature20}. Considering
that the optimization goal is to minimize the average/total weighted
JCT, the output score $q_{v}^{i}$ of node $v$ is multiplied by the
weight $w_{i}$ of the job $i$ to generate a priority score $p_{v}(1\leq v\leq|V|)$.
Finally, a priority list $P=$ $\left(p_{1},p_{2},\ldots,p_{|V|}\right)$
is generated as the scheduling policy, where $p_{v}(1\leq v\leq|V|)$
corresponds to the priority value of schedulable coflows in the execution
of the job DAG, and a higher $p_{v}$ stands for a higher priority.
A formal description of the determination of the priority order is
given in Algorithm 2. 
\begin{algorithm}[tbh]
\caption{Determining priority list}
{\begin{algorithmic}[1] \Require Coflows in the job, job DAGs
and importance weights \Ensure Priority list $P$ of schedulable
coflows \Linecomment TRAINING \State Training based on policy gradient
$\leftarrow$ call Algorithm 4. \State Multiplied Priority score
by job weight \State Return priority list $P=\left(p_{1},p_{2},\ldots,p_{|V|}\right)$
\end{algorithmic} \label{alg:alg2} }
\end{algorithm}

\subsection{Scheduling Active Coflows}

Algorithm 3 is designed to schedule active coflows based on the priority
list. In other words, it is responsible for determining which flows
in active coflows are eligible to occupy network resources to be scheduled.
Due to a lack of network resources, flows in active coflows may be
blocked. Therefore, before scheduling a flow of an active coflow,
we need to check whether the ingress port and the egress port have
free resources. Finally, when a flow is scheduled, we need to release
the occupation of network resources.

\begin{algorithm}[tbh]
\caption{Scheduling active coflows}
{\begin{algorithmic}[1] \Require priority list of schedulable
coflows $P$ \Ensure flow rate of each active coflow \Linecomment
SCHEDULING \For{C in $\mathbb{C}$} \State$r$ = min $\left\{ I_{p},E_{q}\right\} $,
subject to $C$ still need to send flow to port $p$ or receive flow
from port $q$ \For{each unfinished flow $F$ in $C$ } \State
allocate bandwidth of $r$ to $F$ \State update $I_{p}$ and $E_{q}$
by deducting $r$ from them \EndFor \For{each finished flow $F$
in $C$ } \State release bandwidth of $r$ to $F$ \State update
$I_{p}$ and $E_{q}$ by adding $r$ from them \EndFor \EndFor \State
allocate remaining bandwidth equally to all flows \State Return flow
rate of each active coflow \end{algorithmic} \label{alg:alg4} }
\end{algorithm}

\begin{algorithm}[tbh]
\caption{Training based on policy gradient}
{\begin{algorithmic}[1] \Linecomment INITIALIZE \State Initialize
all neural networks \For{each iteration} \State Episode length
$\textit{l}\sim\operatorname{exponential}\left(\textit{l}_{\text{mean }}\right)$
\State Run episodes $\textit{i}$ = 1 to $\textit{N}$ : \State
$\quad\left\{ s_{1}^{i},a_{1}^{i},r_{1}^{i},\ldots,s_{l}^{i},a_{l}^{i},r_{l}^{i}\right\} \sim\pi_{\theta}$
\State Compute total reward: $R_{k}^{i}=\sum_{k^{\prime}=k}^{l}r_{k^{\prime}}^{i}$
\For{$\textit{k}$ = 1 to $\textit{l}$} \State Compute baseline
value: $b_{k}=\frac{1}{N}\sum_{i=1}^{N}R_{k}^{i}$ \For{$\textit{i}$
= 1 to $\textit{N}$} \State $R_{k}^{i}=\sum_{k^{\prime}=k}^{l}r_{k^{\prime}}^{i}$
\EndFor \For{$\textit{i}$ = 1 to $\textit{N}$} \State$\Delta\theta=\Delta\theta+\nabla_{\theta}\log\pi_{\theta}\left(s_{k}^{i},a_{k}^{i}\right)\left(R_{k}^{i}-b_{k}\right)$
\EndFor \EndFor \State$l_{\text{mean}}=l_{\text{mean}}+\epsilon$
\State$\theta=\theta+\alpha\Delta\theta$ \EndFor \end{algorithmic}
\label{alg:alg3} }
\end{algorithm}

\section{Experiment and Evaluation}

\label{sec:experiment and evaluation} In this section, we use the
traces of Facebook \cite{literature27} to test the performance of
the proposed model and provide simulation results and detailed performance
analysis.

\subsection{Simulation Settings}

\textbf{Environment:} For the simulation environment, we create a
online coflow scheduling simulator with Python 3.8 and build neural
networks based on Pytorch. The simulation runs on the Ubuntu 19.04
operating system \cite{literature30}.

\textbf{Workload:} Our workload is generated based on Facebook trace
\cite{literature27}, collected from a 3000-machine, 150-rack MapReduce
cluster at Facebook. This trajectory is widely used in simulation
\cite{literature6,literature12,literature31}, which contains 526
coflows that are scaled down to a 150-port fabric with exact inter-arrival
times. We divide 526 coflows into training set, validation set, and
test set at an approximate ratio of 8:1:1. In each set, randomly select
several coflows, and construct the job DAG by randomly generating
an adjacency matrix. For each coflow, the Facebook trace contains
sender machines, receiver machines, and transmitting bytes at the
receiver level, not the flow level. Thus we partition the bytes in
each receiver to each sender pseudo-uniformly to generate flows. In
addition, the trace only contains the information of each coflow,
not the job DAG. Therefore, to evaluate our algorithms in the scenarios
of dependencies, we randomly select coflows to form the job DAGs so
that each job contains $n$ coflows in expectation and n-1 dependencies
are generated randomly. We randomly selected $P$ machines from the
trace as servers. The arrival time of each job obeys a Poisson distribution
$P(\lambda)$; the weights of the jobs follow a uniform distribution.

\textbf{Metrics:} Here, we evaluate several metrics of our model,
including minimizing the average/total weighted JCT, average JCT.
Minimizing the average JCT is just a special case of minimizing the
total weighted JCT, where all jobs have equal weight.

\textbf{Benchmark algorithms:}

1. Varys \cite{literature6} proposes a Smallest-Effective-Bottleneck-First
(SEBF) heuristic algorithm to greedily schedule coflows based on its
bottleneck flow's completion time.

2. IAOA (Information-Agnostic Online Algorithm) \cite{literature32}
formulates the weighted coflow completion time minimization problem
and proposes a heuristic solution with an approximation factor of
2 to the optimal solution. However, IAOA did not consider the dependency
between coflows in job DAGs

3. DeepWeave \cite{literature14} is the first to use a reinforcement
learning framework to automatically generate efficient coflow scheduling
strategies in job DAGs. However, DeepWeave does not support the goal
of total weighted JCT because it considers the scheduling of jobs
one by one. Therefore, to compare our model with DeepWeave, we first
evaluate our algorithm in a particular case, that is, all jobs have
the same weight, that is, a non-weighted scene. In this case, the
optimization goal is indeed equivalent to minimizing the average JCT.

\textbf{Model Evaluation:}

We randomly select several batches of jobs (one batch contains 1000
jobs) from the test set to generate their job DAGs, and use the Pickle
module in Python to store job data. Under the same input, we compare
the average weighted job completion time of the training model with
the state-of-the-art baseline algorithms.

\subsection{Simulation Results}

\begin{figure}[h]
\centering \includegraphics[width=0.4\textwidth,height=5cm]{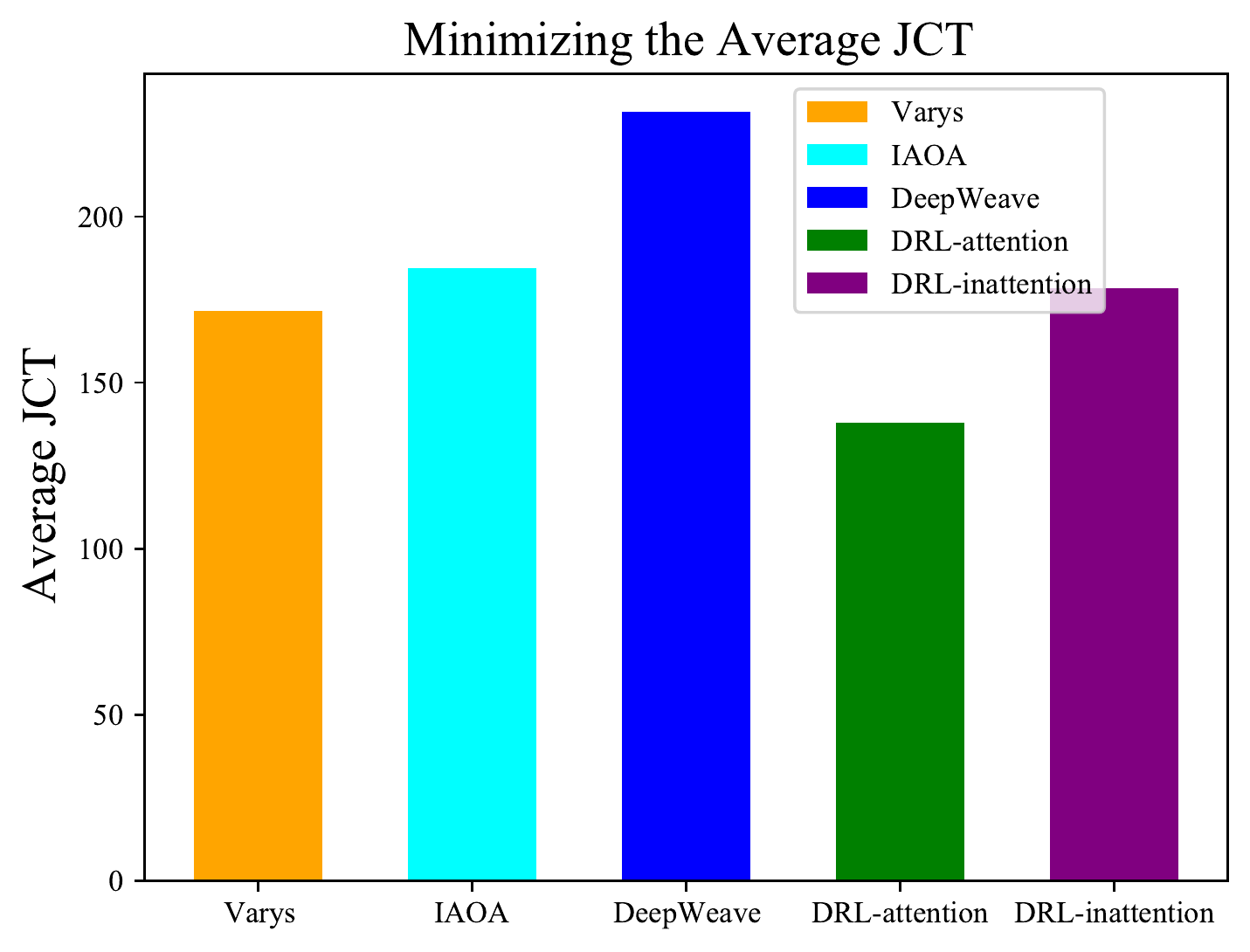}
\caption{Comparison of Average JCT}
\label{fig:single_jct}
\end{figure}

First, for the online scenario where multiple jobs (DAGs) coexist,
we consider how to schedule these jobs with the same weights. In this
case, the optimization objective is equivalent to minimizing the average
job completion time (JCT). We build a job DAG generator to generate
Poisson flow jobs; the time interval of job arrival obeys an exponential
distribution. Unless otherwise specified, we choose the number of
ports $P$ = 20, the average job arrival rate $\lambda$ = 20 and
the average number of coflows $n$ = 8. In the experiment, we compare
our algorithm with DeepWeave, both based on DRL and Varys \cite{literature6}
and IAOA \cite{literature32} that were also used as popular non-ML
baselines for comparison with DeepWeave \cite{literature14}. In addition,
we add an ablation experiment to compare the performance of the retrained
model under the same settings, removing the self-attention layer.
The results of the average JCT is shown in Fig. \ref{fig:single_jct}.
Compared with DeepWeave, the average JCT of our algorithm is reduced
by up to 40.42\%, and the completed jobs is at least 1.68 times more.
Besides, in the case of removing the self-attention layer, the performance
of the model loses 29.36\%.

\begin{figure}[h]
\centering \includegraphics[width=0.4\textwidth,height=5cm]{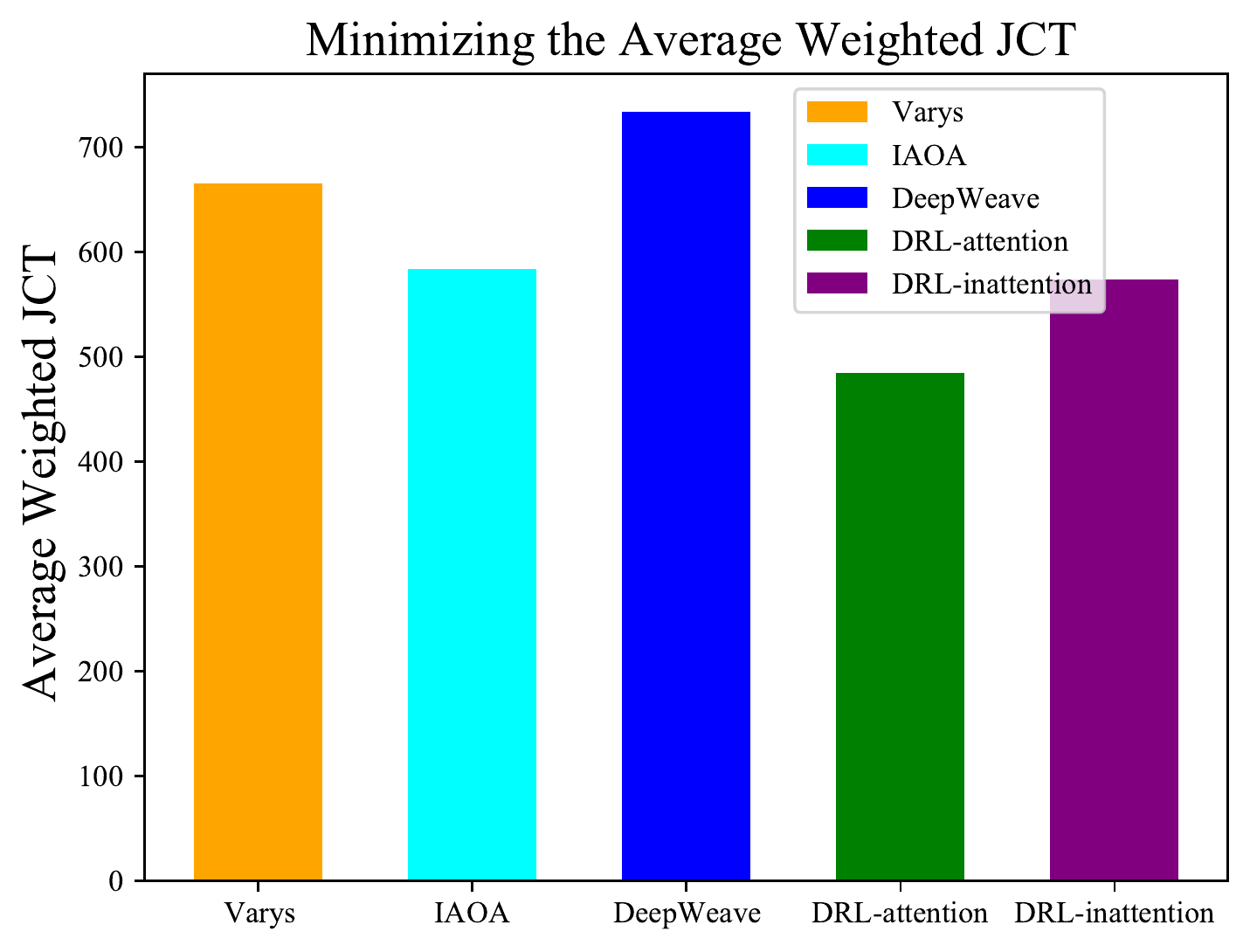}
\caption{Comparison of Average Weighted JCT}
\label{fig:single_weighted}
\end{figure}

Next, we consider the importance of different jobs and minimize the
average weighted job completion time (the average weighted JCT). For
DeepWeave, in the case of online, each arriving job will need to wait
until the previous job is completed before it starts to be released.
Consequently, the importance weight of the job does not affect its
output policy. On contrast, our model can handle the coexistence of
multiple jobs, allowing coflows of different arriving jobs to be scheduled
simultaneously, without waiting for the completion of the previous
job, which can significantly reduce the average JCT. In our model,
the output of the policy network contains the priority scores of schedulable
nodes from different jobs and generate the final priority scheduling
list by multiplying the job's priority score with its weight. Fig.
\ref{fig:single_weighted} shows the performance comparison of different
schemes in the online situation.

\begin{figure}[h]
\centering \includegraphics[width=0.4\textwidth,height=5cm]{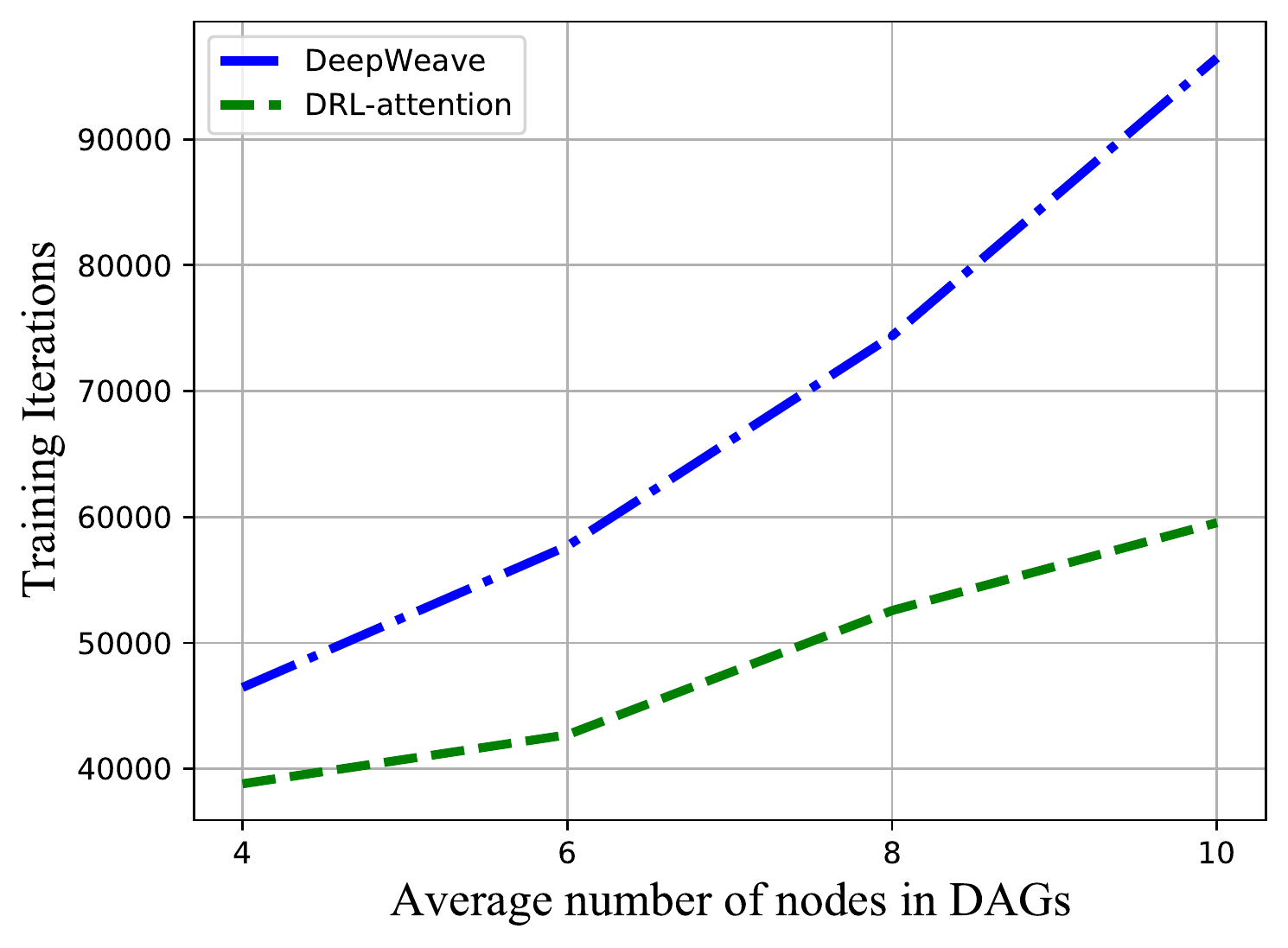}
\caption{Training Iterations vs DAG Size}
\label{fig:node_iterations}
\end{figure}

In order to illustrate better scalability of our proposed model, we
compare the number of iterations required by the proposed model and
DeepWeave as the average number of nodes in the training set increases
in Fig. \ref{fig:node_iterations}. DeepWeave feeds a flat embedding
vector containing the information of the whole job to the policy network,
making the (input) size of the neural network implementing the policy
network proportional to the number of nodes ($n$) in the job DAG.
As $n$ increases, processing a $d$-dimensional feature vector for
the DAG requires a size $O(dn)$ policy network that is difficult
to train. In contrast, in our deep RL model, the policy network scale
is related only to the feature dimension $d$, and hence reduced from
previously $O(dn)$ to $O(d)$.

\begin{figure}[htb]
\centering \includegraphics[width=0.4\textwidth,height=5cm]{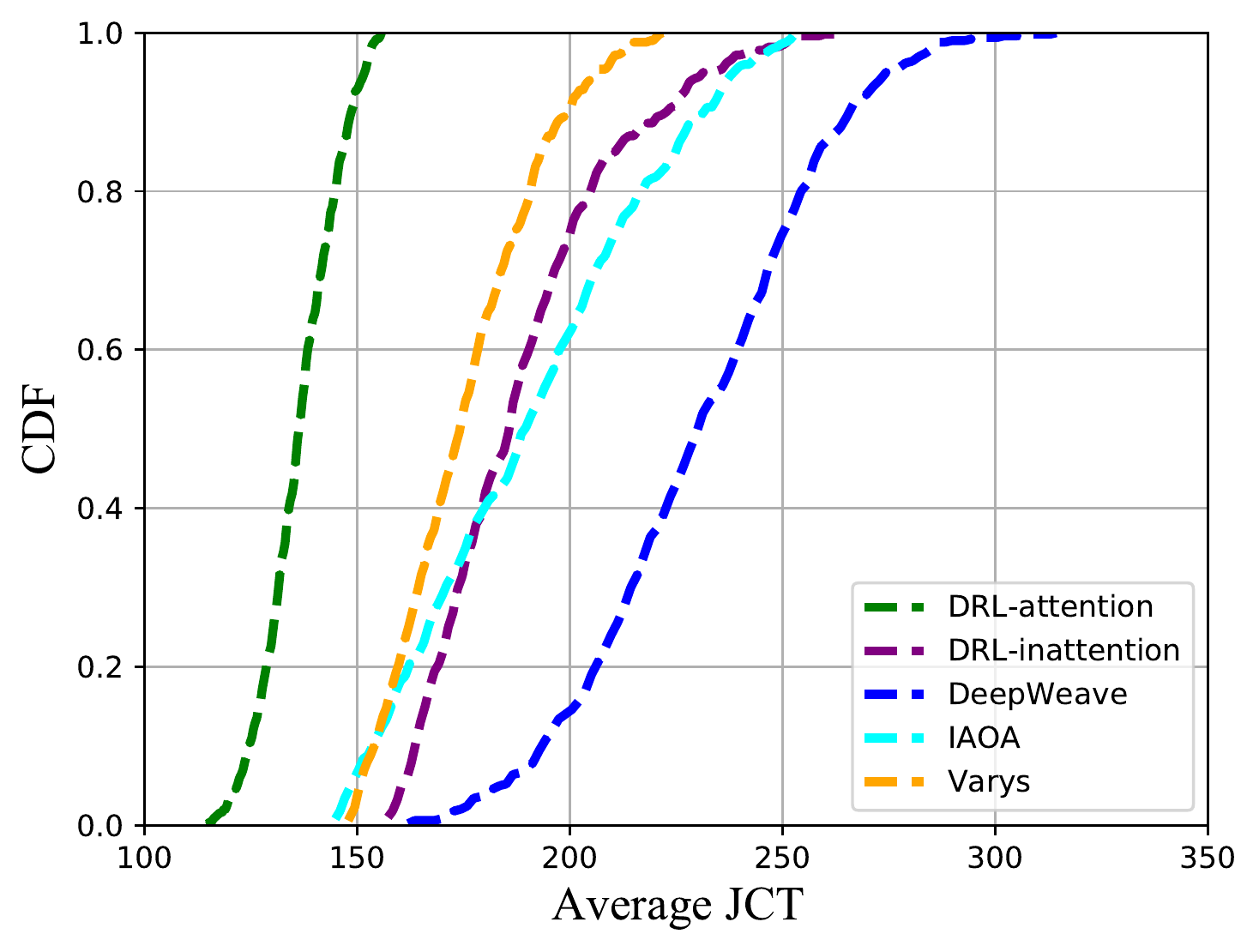}
\caption{CDF of Average JCT}
\label{fig:jct_cdf}
\end{figure}

\begin{figure}[htb]
\centering \includegraphics[width=0.4\textwidth,height=5cm]{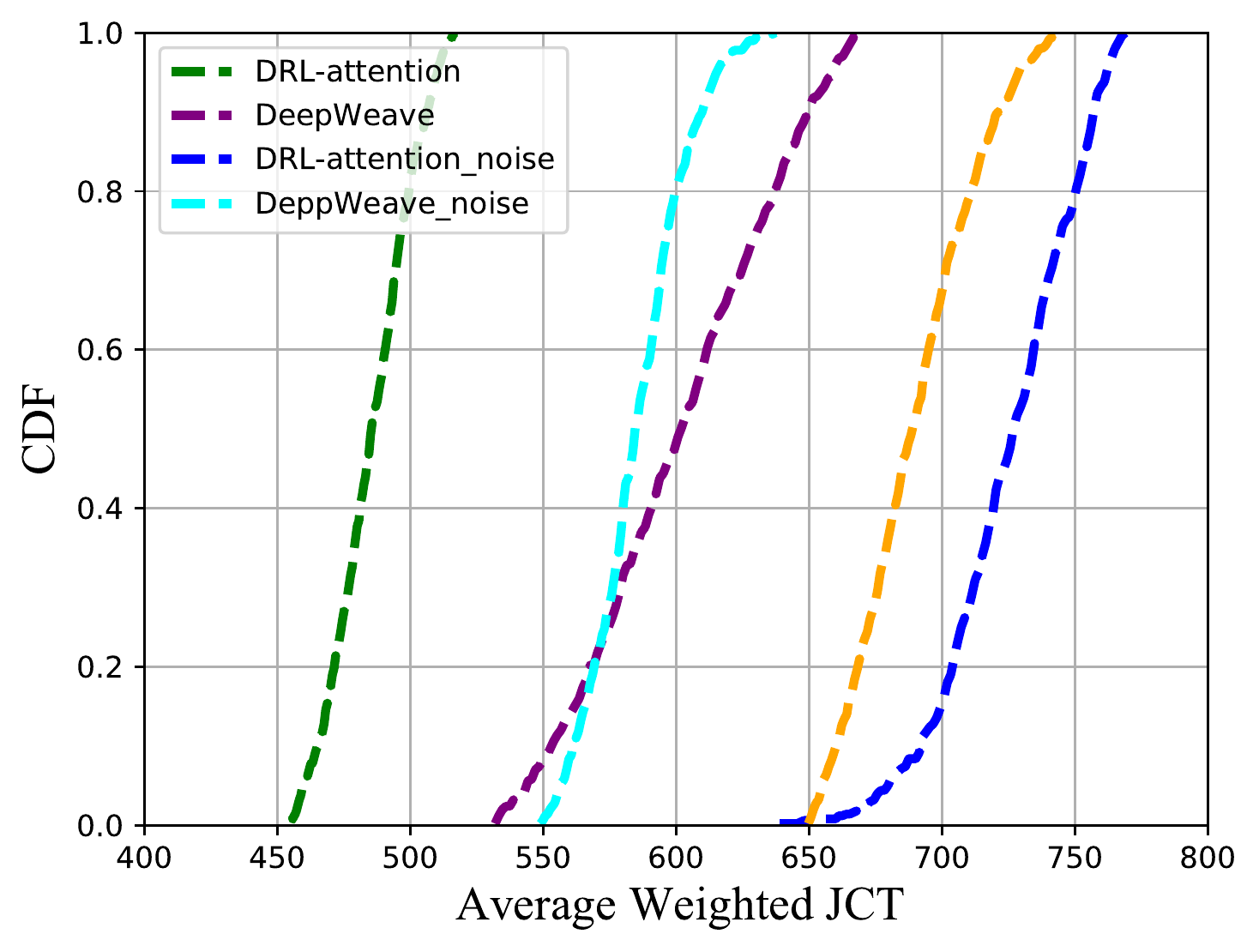}
\caption{CDF of Average Weighted JCT}
\label{fig:weighted_cdf}
\end{figure}

As illustrated, because we generate jobs by randomly combining Facebook
traces and obtaining coflows with random dependencies, it is necessary
to investigate the cumulative distribution function (CDF) of the average
JCT and average weighted JCT. As shown in Fig. \ref{fig:jct_cdf}
and Fig. \ref{fig:weighted_cdf}, the performance and stability of
our model are better than others: for our model, the average JCT varies
from about 113 to 155, and the average weighted JCT varies from about
475 to 536.

The training set and test set of the above experiments are randomly
selected from Facebook trajectories and have the same internal distribution.
However, in a complex data center network, data transmission is often
accompanied by various noises, which are not included in the sampled
data. As a common issue in machine learning models, online coflows
scheduling model based on deep reinforcement learning is sensitive
to noise, yielding significant variance between runs. To explore the
robustness of the proposed model and DeepWeave, we add noise following
normal distribution to the test data set and compare the changes of
the cumulative distribution of job completion time.

\begin{figure}[htb]
\centering \includegraphics[width=0.4\textwidth,height=5cm]{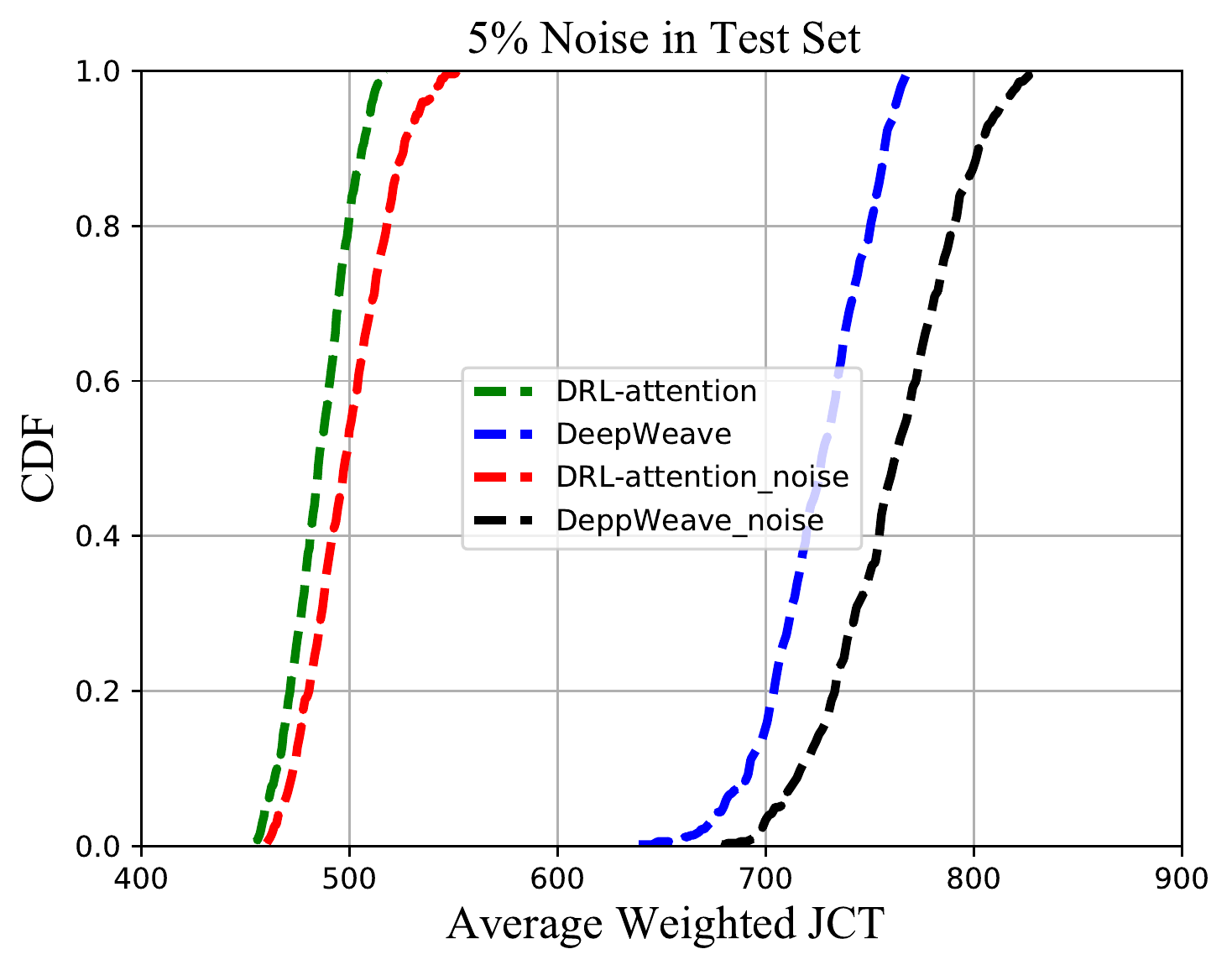}
\caption{CDF of Average Weighted JCT with 5\% Noise}
\label{fig:noise_5}
\end{figure}

\begin{figure}[htb]
\centering \includegraphics[width=0.4\textwidth,height=5cm]{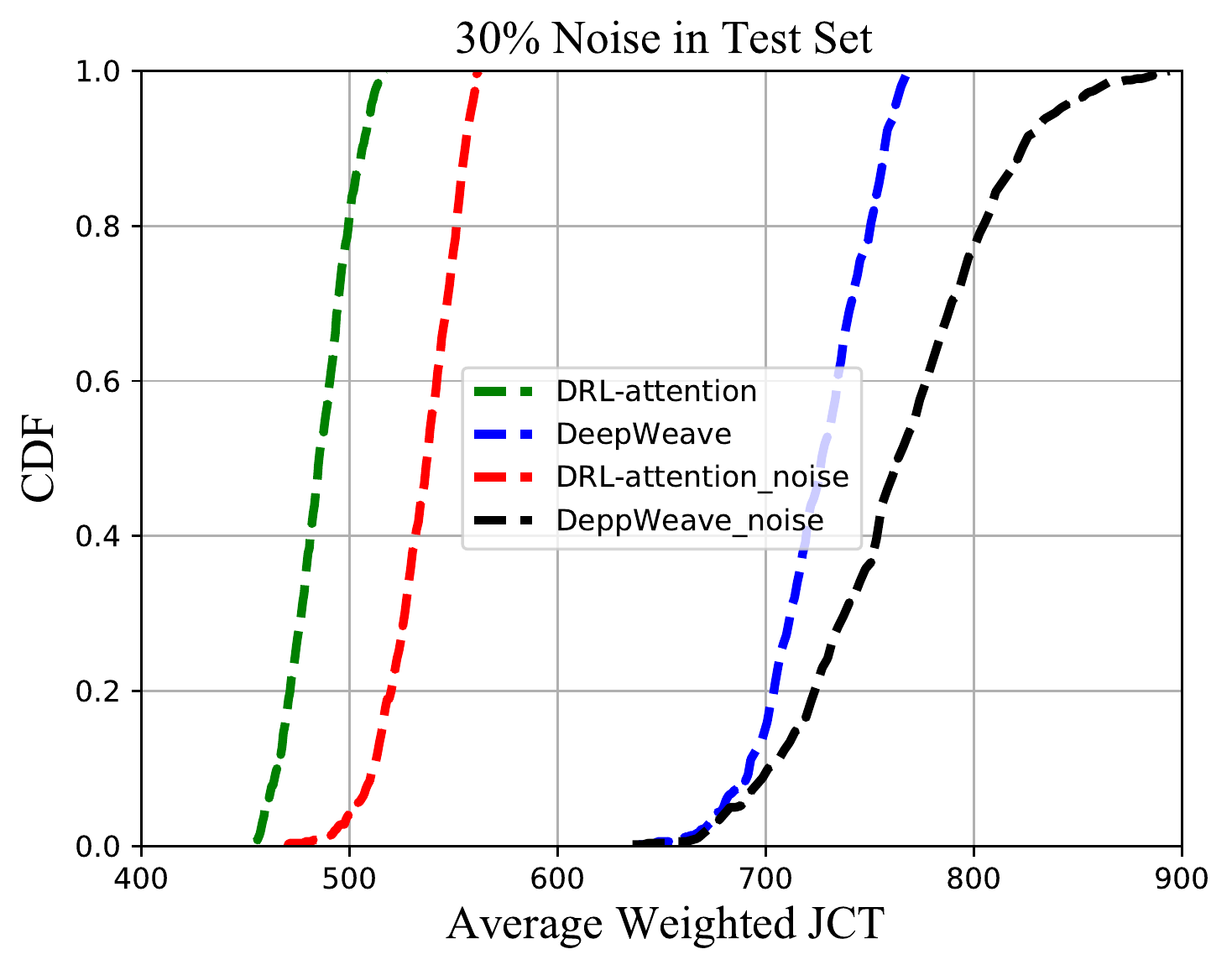}
\caption{CDF of Average Weighted JCT with 30\% Noise}
\label{fig:noise_30}
\end{figure}

When we add a small amount (5\%) of noise, the cumulative distribution
changes of the two models are shown in Fig. \ref{fig:noise_5}. Compared
with DeepWeave \cite{literature14}, the variance and mean differences
on job completion time of the proposed model are smaller. Furthermore,
when the amount of noise in the test set is increased to 30\%, both
distributions of the two models diverge significantly from the original,
as shown in Fig. \ref{fig:noise_30}. DeepWeave has a more significant
variance than our model, because DeepWeave \cite{literature14} requires
to train a large policy network and is prone to over-fitting, resulting
in poor robustness, as the result of directly feeding a high-dimensional
embedding vector containing all the job information as the input of
the policy network.

\begin{figure}[htb]
\centering \includegraphics[width=0.4\textwidth,height=5cm]{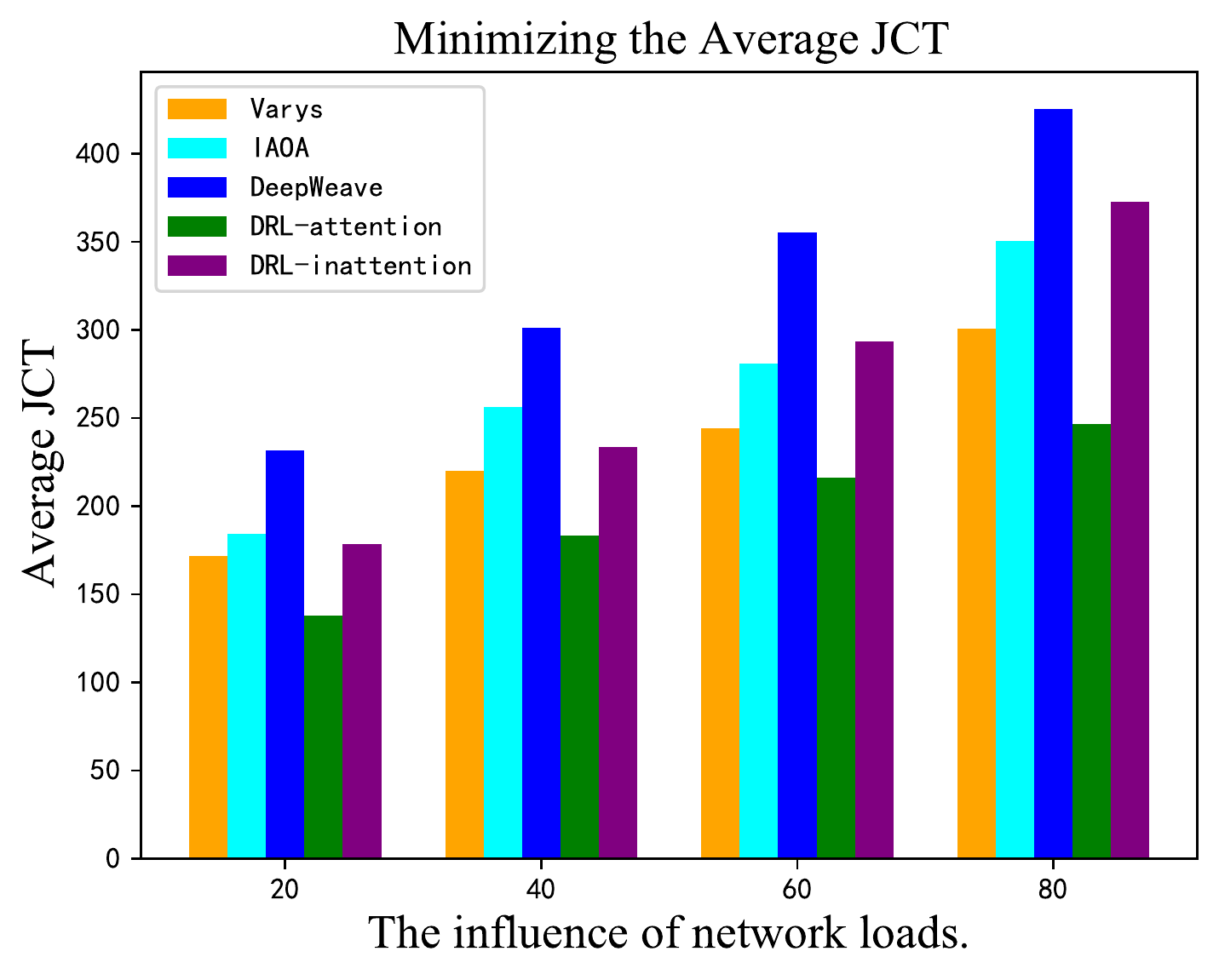}
\caption{Influence of Network Loads on Average JCT}
\label{fig:compare_jct}
\end{figure}

\begin{figure}[!t]
\centering \includegraphics[width=0.4\textwidth,height=5cm]{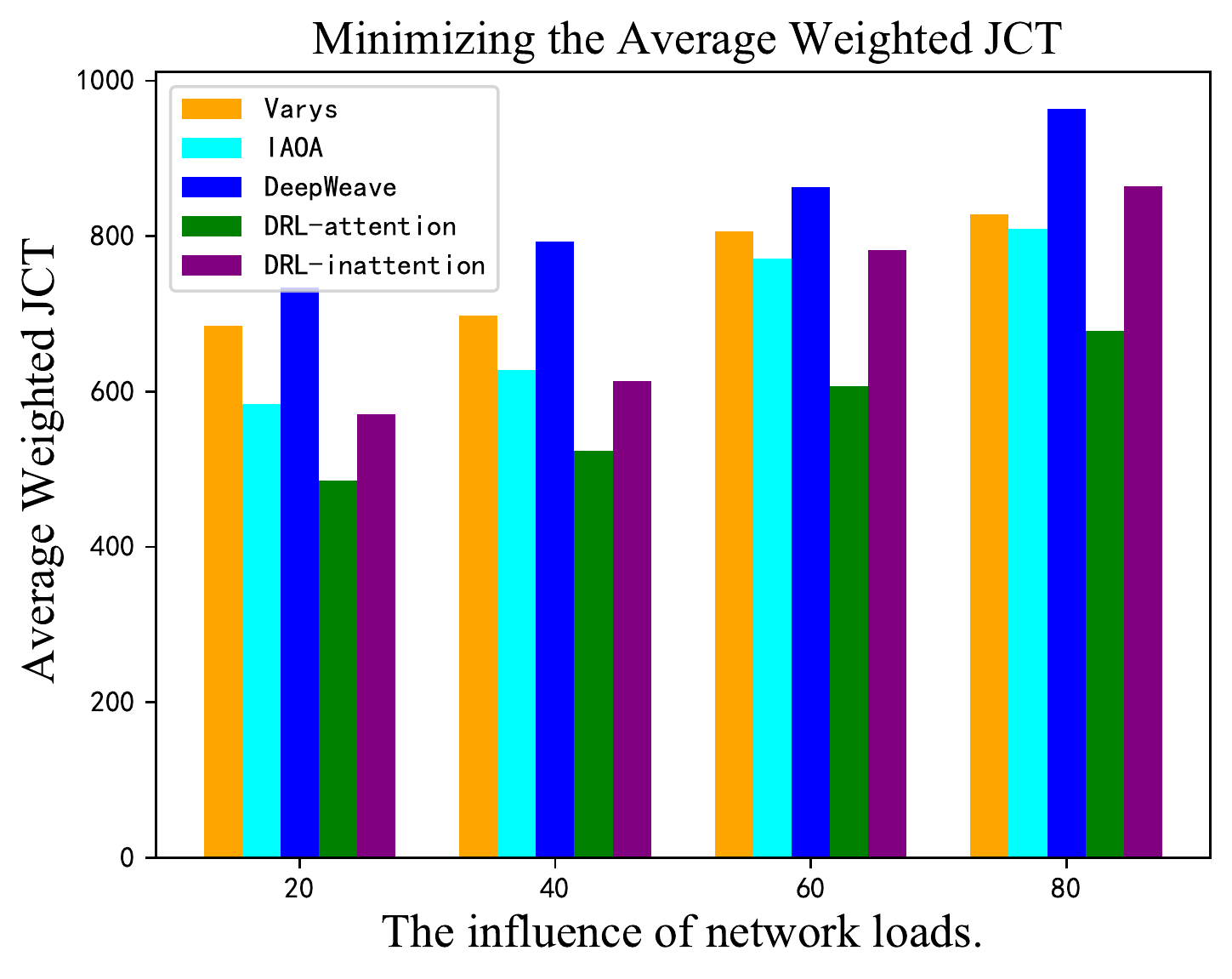}
\caption{Influence of Network Loads on Average Weighted JCT}
\label{fig:compare_weighted}
\end{figure}

Then, we investigate the influence of the average job arrival rate,
i.e. the parameter $\lambda$ in Poisson process, on the average job
completion time (JCT) and average weighted JCT, where a smaller $\lambda$
corresponds to a smaller load, indicating fewer jobs arriving in a
unit time. As shown in Fig. \ref{fig:compare_jct} and Fig. \ref{fig:compare_weighted},
with the increase of network load, there are a large number of job
queues in DeepWeave, resulting in a dramatic rise in both average
JCT and average weight JCT.

\section{Conclusion}

\label{sec:conclusion} In this paper, we proposed an Attention-based
Deep Reinforcement Learning (DRL) Model to generate coflow scheduling
policies for multi-stage jobs with a policy network of significantly
reduced size, which can process job DAGs of arbitrary sizes and shapes
representing jobs with arbitrary coflows while ensuring the accuracy
of the scheduling strategy. Based on this model, we consider the coflow
scheduling problem for online multi-stage jobs, and developed an effective
online algorithm . Our work addresses the main challenge of building
a scalable policy network in applying deep reinforcement learning
to generate coflow scheduling strategies for arbitrary-size job DAGs,
and presents a novel DRL model for online coflow scheduling that is
empowered by the pipelined-DAGNN encoding job DAGs and self-attention
mechanism capturing the interactions among schedulable coflows. Our
model improves the existing work of deploying DRL for coflow scheduling
in scheduling quality for the common goal of minimizing job completion
time, scalability and robustness (sensitivity to noise).

We notice that most existing machine-learning-based coflow scheduling
models, including the state-of-the-art work DeepWeave \cite{literature14}
and our proposed model, have not considered preemptive scheduling,
which allows reordering of the coflows in the active coflow set. The
non-preemption assumption of the active coflows makes these models
more stable and easier to learn effective output strategies. In the
future, we will study DRL based models for preemptive scheduling.
Direct application of our proposed model requires a larger action
space and more frequent decisions to be made by the scheduling agent,
which will increase the training difficulty and reduce the stability
of the model significantly. Consequently, we will investigate the
deployment of multi-agent reinforcement learning techniques \cite{literature33,literature34,literature35}
to design an effective preemptive scheduling model with the strategy
of deploying independent agents to make preemption decisions autonomously.

\section*{Acknowledgement}

This work is supported by Key-Area Research and Development Plan of
Guangdong Province \#2020B010164003. The corresponding author is Hong
Shen.

\bibliographystyle{IEEEtran}
\bibliography{sample-base}

\end{document}